\documentclass[aps,prd,twocolumn,nofootinbib,floatfix]{revtex4-1}
\usepackage{setspace}
\usepackage{amsthm}
\theoremstyle{definition}
\newtheorem*{definition}{Definition}
\usepackage{graphicx}
\usepackage{dcolumn}
\usepackage{multirow}
\usepackage{subfigure}
\usepackage{times,mathptm}
\usepackage{float}
\usepackage{color}
\usepackage{amsmath,amsfonts}
\usepackage{mathptmx}
\usepackage{mathrsfs}
\usepackage{bbm}
\usepackage{bm}
\usepackage{xfrac}

\newcommand{\beq}{\begin{equation}}
\newcommand{\eeq}{\end{equation}} 
\newcommand{\bea}{\begin{eqnarray}}
\newcommand{\eea}{\end{eqnarray}}

\newcommand{\E}{\mathcal{E}}
\newcommand{\V}{\mathcal{V}}

\renewcommand{\d}{\delta}

\renewcommand{\l}{\lambda}

\newcommand{\p}{\phi}
\renewcommand{\b}{\beta}
\renewcommand{\a}{\alpha}

\newcommand{\tr}{\text{Tr}}

\newcommand{\vx}{{\vec{x}}}
\newcommand{\vy}{{\vec{y}}}
\newcommand{\vz}{\vec{z}}

\newcommand{\n}{\nu}
\newcommand{\m}{\mu}

\newcommand{\g}{\gamma}
\renewcommand{\r}{\rho}
\newcommand{\e}{\epsilon}
\newcommand{\s}{\sigma}

\newcommand{\N}{{\cal N}}

\newcommand{\vph}{\varphi}
\newcommand{\oh}{\frac{1}{2}}

\newcommand{\dg}{\dagger}
\newcommand{\non}{\nonumber}

\newcommand{\rf}[1]{(\ref{#1})}
\newcommand{\ra}{\rightarrow}
\newcommand{\pa}{\partial}
\renewcommand{\vec}[1]{\bm #1}

\usepackage{ulem}

\begin{document}

\title{A confinement criterion for gauge theories with matter fields} 

\bigskip
\bigskip

\author{Jeff Greensite and Kazue Matsuyama}
\affiliation{Physics and Astronomy Department, San Francisco State
University,   \\ San Francisco, CA~94132, USA}
\bigskip
\date{\today}
\vspace{60pt}
\begin{abstract}

\singlespacing
 
   A generalization of the Wilson loop area-law criterion is proposed, which is applicable to gauge theories with
matter in the fundamental representation of the gauge group.   This new criterion, like the area law, is stronger than
the statement that asymptotic particle states are massive color singlets, which holds even for theories described by the
Brout-Englert-Higgs mechanism.    

\end{abstract}

\maketitle

\singlespacing
\section{\label{intro}Introduction}

   It is hardly news that QCD is confining, but the meaning of the word ``confinement'' as it applies to QCD, or to any gauge theory with matter fields that break global center symmetry, is surprisingly subtle.  Probably what most people have in mind is ``color confinement,'' meaning that the spectrum only contains particles which are color singlets. We will refer to this property as ``C-confinement.'' But if this is what one means by confinement, then it must be recognized that a gauge-Higgs theory, in which the gauge symmetry is completely broken by Higgs fields in the fundamental representation, is also confining in this sense.\footnote{The suggestion that a local gauge symmetry can be spontaneously broken is rather misleading, in view of Elitzur's theorem \cite{Elitzur:1975im}, but the terminology is standard in almost all textbooks on quantum field theory and it seems hopeless at this stage to fight tradition.}   Such theories are much like the weak interactions.  There are heavy vector bosons (the ``W''-particles) which acquire a mass via the Brout-Englert-Higgs mechanism, and there are only Yukawa forces between charged particles.  It was realized long ago by Fradkin and Shenker \cite{Fradkin:1978dv}, building on
a theorem due to Osterwalder and Seiler
\cite{Osterwalder:1977pc}, that the spectrum of gauge-Higgs theories of this type also consists only of color singlets.  The argument is
based on the fact that in a lattice formulation of gauge-Higgs theories the Higgs-like region of the phase diagram in coupling-constant space is 
connected to the confinement-like region.  In the confinement-like region there is flux tube formation and a linear potential between heavy sources over some distance range, followed by string breaking, just as in QCD.  In this region we clearly have, as in QCD, a spectrum which consists of color singlets.  However, as shown in the work of Fradkin, Shenker,
Osterwalder and Seiler (FSOS) \cite{Fradkin:1978dv,Osterwalder:1977pc}, there is always a path in coupling-constant space between points in the confinement-like region to points in the Higgs-like region such that there are no discontinuities in local gauge-invariant quantities anywhere along the path.\footnote{Although local gauge symmetries cannot be spontaneously broken, according to Elitzur's theorem, global symmetries left over after gauge fixing {\it can} break spontaneously, so one might think that the distinction between a confining and non-confining theories might be related to the breaking of such remnant gauge symmetries.  The problem with that idea is
that different remnant symmetries associated with different choices of gauge fixing break in different places in the phase
diagram \cite{Caudy:2007sf},  which makes such proposals appear rather arbitrary and gauge-dependent.}  This implies, in particular, continuity of the spectrum along the path, and this in turn means that there can be no
sudden change from a color singlet to a color non-singlet spectrum.   Asymptotic particle states such as the physical $W$ particles are created, in the Higgs-like region, by gauge-invariant composite operators \cite{tHooft:1979yoe,Frohlich:1981yi,*Frohlich:1980gj}.  The appropriate formalism for dealing with such states 
(and for re-deriving the perturbative results of the standard treatment) was
developed by Fr\"ohlich, Morchio, and Strocchi \cite{Frohlich:1981yi,*Frohlich:1980gj}, and has been further advanced by Maas et al.\  \cite{Egger:2017tkd,*Torek:2016ede}.  Thus, if by confinement we mean a spectrum of color singlets, then the weak interactions of a gauge-Higgs theory, and even the charge-screening dynamics of
an ordinary electric superconductor, are also confining.   
In this respect the term ``color confinement'' seems somehow unsatisfactory, if we think of confinement as associated, e.g., with color electric flux tube formation, linear Regge trajectories, and a linear static quark potential.  

   On the other hand, apropos the linear potential, there is a different and stronger meaning that one can assign to the word ``confinement'' in a pure SU(N) gauge theory, a meaning which goes beyond the statement that all asymptotic particle states are color neutral.   A pure SU(N) gauge theory is of course C-confining, with a spectrum of color-singlet glueballs and resonances, but such theories {\it also} have the property that the static quark potential rises linearly or, equivalently, that planar Wilson loops have an area-law falloff.  This property is not equivalent
to C-confinement, nor is it equivalent to a mass gap, as the gauge-Higgs example makes clear.  This raises an interesting question: is there any way to generalize the  Wilson area law criterion to gauge theories with string-breaking, and matter in the fundamental representation?

\section{Separation of charge Confinement}

\subsection{S${}_c$-confinement, pure gauge theories}

    Consider a rectangular $R \times T$ Wilson loop in Euclidean spacetime.  The Wilson loop area law criterion can be reformulated in the following way.  Begin with a state at Euclidean time $t=0$
\beq
          \Psi_{\overline{q} q}(t=0) \equiv \overline{q}^a(\vx) V_0^{ab}(\vx,\vy;A) q^b(\vy) \Psi_0 \ ,
\label{extra1}
\eeq
where the $\overline{q}, q$ operators create an extremely massive static quark-antiquark pair with separation $R=|\vx-\vy|$, $\Psi_0$ is the vacuum state, and
\beq
    V_0(\vx,\vy;A) = P\exp[ig\int_\vx^\vy dz^\m A_\m(z)]
\label{Wline}
\eeq
is a Wilson line joining points $\vx,\vy$ at time $t=0$ along a straight-line path.  Let this state evolve in Euclidean time.
Since the quarks are static and the initial state is gauge invariant, the state at Euclidean time $t$ has the form
\beq
 \Psi_{\overline{q} q}(t) = \overline{q}^a(\vx) V_t^{ab}(\vx,\vy;A) q^b(\vy) \Psi_0 \ ,
\label{extra2}
\eeq
where $V_t^{ab}(\vx,\vy;A)$ is a gauge bi-covariant operator which transforms, under a local gauge transformation $g(\vx,t)$, as
\beq
          V_t^{ab}(\vx,\vy;A) \ra V'^{ab}_t(\vx,\vy;A) = g^{ac}(\vx,t) V_t^{cd}(\vx,\vy;A) g^{\dg db}(\vy,t) \ .
\label{bicovariant}
\eeq
Since the evolution is in Euclidean space, the energy expectation value of $\Psi_{\overline{q} q}(t)$ decreases monatonically,  converging in the $t \ra \infty$ limit to the ground state energy $E_0(R)$ of a static quark-antiquark pair.\footnote{In order that the self-energy contribution is finite, a lattice regularization is implicitly assumed.}  Then it is easy to see that the area-law falloff of a large Wilson loop in pure gauge theory is equivalent to the following property, which we will call ``separation-of-charge confinement," or
simply ``S${}_c$-confinement'':\footnote{The still shorter abbreviation ``S-confinement'' (with ``S" for "separation") used in an earlier version of this article would be preferable, but that terminology (with ``S" for ``smooth") has already been used in the literature (c.f.\ \cite{Csaki:1996sm}) in another context. We thank Erich Poppitz for bringing this point to our attention.}  \\

\theoremstyle{definition}
\begin{definition}{}
   Let $V(\vx,\vy;A)$ be a gauge bi-covariant operator transforming as in \rf{bicovariant}, and let $E_V(R)$, with $R=|\vx-\vy|$ be the 
energy of the corresponding state
\beq
        \Psi_V \equiv  \overline{q}^a(\vx) V^{ab}(\vx,\vy;A) q^b(\vy) \Psi_0
\label{Vstate}
\eeq
above the vacuum energy $\E_{vac}$.
\textbf{\textit{S${}_c$-confinement}} means that there exists an asymptotically linear function $E_0(R)$, i.e.\
\beq
    \lim_{R\ra \infty} {dE_0 \over dR} = \s >0 \ ,
\eeq
such that
\beq
        E_V(R) \ge E_0(R) 
\label{criterion}
\eeq
for any choice of bi-covariant $V(\vx,\vy;A)$. \\
\end{definition}

For a pure gauge theory this means that the energy $E_V(R)$ of any quark-antiquark state is bounded from below by a potential which
grows like $\s R$ at large $R$.  If the ground state saturates the bound \rf{criterion}, then S${}_c$-confinement implies that a Wilson loop will have an area law falloff in the large $R,T$ limit with the coefficient of the area equal to $\s$. \\

\subsection{S${}_c$-confinement, gauge + matter theories}

    We now propose that S${}_c$-confinement should also be regarded as the confinement criterion for gauge theories with matter fields, and
it is a much stronger criterion than the C-confinement criterion defined earlier.  The crucial element in the criterion, for gauge + matter theories such as QCD and gauge-Higgs theory, is that the
bi-covariant operators $V^{ab}(\vx,\vy;A)$ {\it must depend only on the gauge field $A$} at a fixed time, and not on the matter fields.  This means that
the asymptotically linear function $E_0(R)$ is not the minimal energy of a static quark-antiquark system, because with (bosonic or fermionic) matter $\phi$ in the fundamental representation of SU(N) one can always construct a covariant operator $Q^a(\vx;A,\phi)$, and use this operator to produce a quark-antiquark state which factorizes into two color singlets, i.e.
\beq
    \Psi_{\overline{q} q} = \{ \overline{q}^a(\vx) Q^a(\vx;A,\p)\} \times \{ Q^{\dg b}(\vy)q^b(\vy;A,\p) \}\Psi_0 \ .
\label{factor}
\eeq
A trivial example is $Q^a(\vx;A,\phi)=\phi^a(\vx)$.
For large separations, the energy expectation value of such factorizable states may have little dependence on $R$.  These states may represent, for example, two non-interacting bound states of a static quark + matter pair, and a static antiquark + matter pair.  States of this kind are excluded in the definition of S${}_c$-confinement, because $V(\vx,\vy;A)$ is not allowed to depend on the matter fields.

   To appreciate the physical meaning of S${}_c$-confinement, and the distinction between S${}_c$- and C-confinement, it useful to recall the origins of both string theory and QCD, and in particular the role played by two experimental facts:  the absence of free quarks, and the existence of linear Regge trajectories.  The C-confinement criterion derives from the absence of free quarks, and the absence of any long-range color field by which a non-zero color charge could be detected.  But as already noted repeatedly, this criterion applies just as well to a gauge-Higgs theory with only Yukawa forces.  S${}_c$-confinement, on the other hand, is most likely related to linear Regge trajectories, and their interpretation in terms of the formation of (metastable) color electric flux tubes.  It is true that in QCD two color charges cannot be widely separated for
very long.  String breaking prevents this.  But there is nothing to prevent us from considering the energy expectation value of physical states such as $\Psi_V$ in eq.\ \rf{Vstate}, which {\it can} have a large separation of matter field color charges, and indeed the highly excited resonances on linear Regge trajectories, or the state of a quark and antiquark which are very suddenly separated from one another, presumably have such a form prior to decay via string breaking.  The S${}_c$-confinement property requires that states of this kind in which, prior to string breaking, two colored particles are widely separated, have an energy which is bounded from below by a linear potential.  Thus the S${}_c$-confinement criterion proposed here is related to the existence of resonances lying on linear Regge trajectories.  

    Suppose QCD were {\it not} S${}_c$-confining, with $E_0(R)$ in \rf{criterion} going to some constant $M$ in the $R \ra \infty$ limit, and
we will denote the mass of a heavy quark-light quark bound state as $M_0$.  It is interesting to consider the consequences.  First imagine that $M < 2M_0$.  In that case isolated quarks and antiquarks at large separations could exist as stable states; we do not even have
C-confinement.  The alternative possibility is $M > 2M_0$.  In that case the quark-antiquark state at large color charge separation would
decay into heavy-light bound states, and possibly other mesons as well.  In other words, the system with large color charge separation is
unstable, and the minimal energy configuration of this type would be a resonance of some kind.  Asymptotic particle states would consist of color singlets, and we still have C-confinement.  On the other hand, the absence of S${}_c$-confinement would be evident in scattering amplitudes.  Replacing the heavy quark sources by light quarks, and assuming $M > 2m_\pi$, we see that the loss of S${}_c$-confinement would change the spectrum of resonances.  Without flux tube formation and the linear relation between quark separation and energy,
there is no reason that excited light quark states should fall on linear Regge trajectories, and the hadronic S-matrix would be quite different from what is observed.   For this reason, we conjecture that QCD is an S${}_c$-confining theory.

\section{Tests of S${}_c$-confinement}

   It is difficult to verify S${}_c$-confinement numerically.  The problem is that the set of all bi-covariant operators $V(\vx,\vy;A)$ is infinite. 
Members of that set include
Wilson lines as in eq.\ \rf{Wline}, or a superposition of Wilson lines on (equal times) curves $C_{\vx \vy}$ between point $\vx, \vy$
\beq
         V(\vx,\vy;A) = \sum_{C_{\vx \vy}} a(C_{\vx \vy})  P\exp\left[i\int_\vx^\vy dz^\m A_\m(z)\right] \ ,
\label{super}
\eeq
or a covariant Green's function such as
\beq
        V(\vx,\vy;A) = \left( {1 \over -D_i D_i} \right)_{\vx \vy}  \ ,
\eeq
where $D_i$ is a covariant derivative.
Showing that one or several of these operators obeys the S${}_c$-confinement criterion is not a proof that all such operators obey the criterion, and this makes a proof S${}_c$-confinement, or even a numerical verification of the property, a challenging exercise.  Perhaps the best indication of S${}_c$-confinement, in a theory with light quarks, would be to demonstrate (as far as possible) that there is a spectrum of resonances lying  on linear Regge trajectories.

  On the other hand, if one can show that a single operator violates the criterion, then that is sufficient to show that the theory is {\it not}
S${}_c$-confining. The question is which operator or operators would be useful in a numerical investigation of this kind.     Certainly the Wilson line operator \rf{Wline} is not helpful.  The energy of the corresponding state $\Psi_V$ would grow linearly with $R$ even in an abelian gauge theory in D=4 dimensions, which is surely non-confining.  The reason is that the Wilson line operator creates a collimated line of electric flux, essentially regardless of the state it operates on, with thickness on the order of the short distance regulator. In this article we will consider two other possibilities.

\subsection{The Dirac state}

   The abelian theory is a useful guide to a choice of $V$, because for the pure abelian theory we actually know the lowest energy eigenstate of the theory containing a static pair of oppositely charged particles.  In this case the $V$ operator is factorizable, and the ground state is the one introduced by Dirac \cite{Dirac:1955uv}:
\beq
    \Psi_{\overline{q} q} = \{ \overline{q}(\vx) G_C^\dg(\vx;A)\} \times \{ G_C(\vy;A)q(\vy) \}\Psi_0 \ ,
\label{abelian}
\eeq
where
\beq
    G_C(\vx;A) = \exp\left[-i \int d^3z ~ A_i(z) \pa_i {1\over 4\pi |\vx-\vz|} \right] \ .
\eeq
It is straightforward to check that this gauge-invariant state is an eigenstate of the Hamiltonian operator in a physical gauge, such as
temporal or Coulomb gauge.  It is equally straightforward to check that $G_C(\vx;A)$ is the gauge transformation which takes an
arbitrary $A$-field into Coulomb gauge, so that in Coulomb gauge the state \rf{abelian} is just
\beq
    \Psi_{\overline{q} q} =  \overline{q}(\vx)  q(\vy) \Psi_0 \ .
\label{coulomb}
\eeq
This abelian example suggests that it may be worthwhile to check, in non-abelian gauge + matter theories, whether the ``Dirac
state'' 
\beq
\Psi_V =   \overline{q}^a(\vx) G_C^{\dg ac}(\vx;A) G_C^{cb}(\vy;A) q^b(\vy) \Psi_0
\label{Psifact}
\eeq
formed from the operator
\beq
          V^{ab}(x,y;A) = G_C^{\dg ac}(\vx;A) G_C^{cb}(\vy;A)
\label{VCoul}
\eeq
satisfies the S${}_c$-confinement criterion, where $G_C(\vx;A)$ is the gauge transformation which takes a non-abelian gauge field $A$ to Coulomb gauge.   In Coulomb gauge this state has the deceptively local appearance
\beq 
            \Psi_V = \overline{q}^a(\vx)  q^a(\vy) \Psi_0 \ ,
\label{cstate}
\eeq
and we want to compute the energy expectation value
\beq
            E_V(R) =  -\lim_{t=0} {d\over dt} \log \left[  \langle \Psi_V| e^{-Ht} |\Psi_V \rangle \over 
                     \langle \Psi_V|\Psi_V \rangle \right]  - \E_{vac} \ .
\eeq
With massive static quarks, $N$ colors, and a lattice regularization, the desired observable is
\bea
    E_V(R) =  -  \log \big\langle {1\over N}\tr[U_0({\bf 0},0) U_0^\dg({\bf R},0)] \big\rangle \ ,
\label{Ecoul}
\eea
evaluated in Coulomb gauge.  Note that in Coulomb gauge there is still a remnant gauge symmetry under constant (i.e.\ space-independent) gauge transformations $g(\vx)=g$. The operator $\overline{q}^a(\vx)  q^a(\vy)$ is invariant under such transformations, while
the individual operator $q^a(\vx)$ is not. 

     In contrast to the abelian theory, it is very difficult to construct $G_C(\vx;A)$ explicitly in terms of the $A$-field in a non-abelian theory, in part because of the Gribov ambiguity.
If, however, we remove this ambiguity (up to the remnant global symmetry) by, e.g., imposing a further condition such as restriction to the fundamental modular region, then there is no doubt that the transformation at least exists.  The fundamental modular region (or ``minimal'' Coulomb gauge), which minimizes the volume average of $\tr A_i A_i$, is difficult to impose numerically, but in principle it is relatively easy to remove the Gribov ambiguity on the lattice in the following way:  First transform the lattice configuration to a completely fixed gauge of some kind, e.g.\ completely fixed axial or Laplacian gauge.  Then
fix to Coulomb gauge by some deterministic over-relaxation algorithm implemented numerically.  This ensures that any two gauge-equivalent lattice configurations will be fixed to a unique Gribov copy, up to a global (i.e.\ position independent) gauge transformation in Coulomb gauge, and we note that \rf{cstate} is invariant under the remnant global symmetry.  The algorithm is also a constructive procedure for generating, non-perturbatively, the gauge transformation $G_C(\vx;A)$.  In practice, however, we do not believe that Gribov copies are an important issue with respect to the confinement problem, since most observables (e.g.\ gauge-dependent correlators of various kinds) generally vary only a little when the gauge fixing algorithm on the lattice is ``improved'' in some way, e.g.\ in an effort to approach the minimal Coulomb gauge.  We will therefore dispense, in the numerical evaluation of $E_V(R)$, with the initial step of fixing to a completely fixed axial gauge.  

\subsection{Pseudo-matter fields}

   Although matter fields are excluded by definition in the bicovariant operator $V(\vx,\vy;A)$, it is nonetheless possible to
construct operators depending only on the gauge field, which transform like matter fields in the fundamental representation.
We will call these operators ``pseudo-matter'' fields, because, while they transform like matter fields, they have no influence
on the probability distribution of the gauge fields.  An example is any eigenstate $\vph^a_n(\vx)$ 
\beq
           (-D_i D_i)^{ab}_{\vx \vy} \vph^b_n(\vy) = \l_n \vph^a_n(\vx) 
\eeq
of the lattice Laplacian operator 
\bea
 \lefteqn{(-D_i D_i)^{ab}_{\vx \vy} = } \non \\
    &=& \sum_{k=1}^3 \left[2 \d^{ab} \d_{\vx \vy} - U_k^{ab}(\vx) \d_{\vy,\vx+\hat{k}}    - U_k^{\dg ab}(\vx-\hat{k}) \d_{\vy,\vx-\hat{k}} 
    \right] \ .  \non \\ 
\eea
Obviously the choice of pseudo-matter field is limitless; any eigenstate of any covariant operator $C^{ab}_{\vx \vy}$ is a field
which transforms in the fundamental representation of the gauge group.  Below we will consider, just as an example, $V(\vx,\vy;A)$ built from the pseudo-matter field $\vph^a_1(\vx)$ corresponding to the smallest eigenvalue of the lattice Laplacian $-D_i D_i$, i.e.
\beq
           V^{ab}(\vx,\vy;A) = \vph_1^a(\vx) \vph_1^{\dg b}(\vy) \ .
\label{pseudo}
\eeq

   If we would make the same construction \rf{pseudo} using dynamical matter fields $\phi(\vx)$ rather than pseudomatter fields, then the corresponding energy expectation value goes to a constant as quark-antiquark separation increases, as discussed below \rf{factor}.  It is interesting to check that this is {\it not} what happens using pseudo-matter fields, in theories that we have argued should be S${}_c$-confining.   Taking the eigenstate $\vph^a_1(\vx)$ of the covariant Laplacian as the pseudo matter field, the lattice expression for the energy expectation value is
\begin{widetext}
\beq
     E_V(R) = -\log \left[ {\langle \{ \vph_1^\dg((\vx,t ) U_0(\vx,t) \vph_1(\vx,t+1) \} \{\vph^\dg_1(\vy,t+1) U_0^\dg(\vy,t) \vph_1(\vy,t) \}\rangle
                              \over \langle \{ \vph_1^\dg((\vx,t ) \vph_1(\vx,t) \} \{ (\vph^\dg_1(\vy,t)  \vph_1(\vy,t) \} \rangle } \right] \ ,
\label{Epm}
\eeq
\end{widetext}
which may be used as a second probe of S${}_c$-confinement in gauge-Higgs theory.

\section{Gauge-Higgs theory} 

\begin{figure}[t]
\centerline{\rotatebox{270}{\includegraphics[width=5.5truecm]{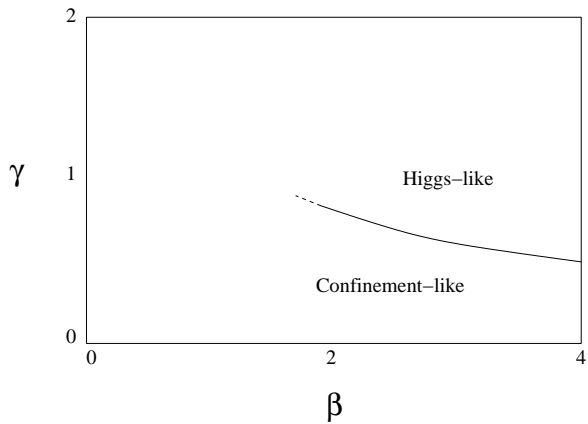}}}
\caption{Schematic phase diagram of the SU(2) gauge-Higgs system.  
The solid line is a line of rapid crossover or weak first-order transitions.} 
\label{fradshenk}
\end{figure}

   Here, as in ref.\ \cite{Caudy:2007sf}, we consider a gauge-Higgs theory on the lattice, 
with a fixed-modulus Higgs field in the fundamental color representation.  For the 
SU(2) gauge group, the Lagrangian can be written in the form \cite{Lang:1981qg}
\bea
     S = \b \sum_{plaq} \oh \mbox{Tr}[UUU^\dg U^\dg] 
       + \gamma \sum_{x,\m} \oh \mbox{Tr}[\phi^\dg(x) U_\m(x) 
\phi(x+\widehat{\m})] \ , \non \\
\label{ghiggs}
\eea
with $\phi$ an SU(2) group-valued field.   We may be confident, because of the work of
FSOS \cite{Fradkin:1978dv,Osterwalder:1977pc}, together with numerical simulations that
have ruled out a massless phase \cite{Campos:1997dc,*Langguth:1985dr,*Jersak:1985nf}, \cite{Bonati:2009pf},
that this theory is C-confining everywhere in the
$\b-\g$ coupling-constant plane, and the phase diagram in this plane looks something like
Fig.\ \ref{fradshenk}, where the line represents a line of 1st order transitions, or rapid crossover.   We have reason to
also believe that the theory is S${}_c$-confining in at least part of the $\b-\g$ plane, at sufficiently small $\g$.
The reason is simply the similarity of gauge-Higgs theory to QCD in this region.  We have already argued
that QCD is S${}_c$-confining: if this were not the case, then it would be possible to separate massive
colored objects far apart from one another, without a proportional cost in energy.  While there is
no proof, gauge-Higgs theory is similar in so many respects to QCD, particularly in the existence 
of color-electric flux tube formation followed by string-breaking, that it would be surprising if
QCD had the S${}_c$-confinement property while gauge-Higgs in the confinement-like region did not.

\subsection{The Dirac operator transition}

    We will now show that the S${}_c$-confinement property does not hold in the Higgs region
of the phase diagram.  Assuming S${}_c$-confinement exists in (and, for that matter, defines) the confinement-like region, this implies
the existence of a transition from S${}_c$-confinement to a region which does not have this property.
As we know from FSOS, this transition from S${}_c$-confinement to only C-confinement cannot be associated 
with any non-analyticity in the free energy or spectrum.  Yet we must insist that the separation of massive color charges
at long distances with or without a proportional cost in energy is a physically meaningful distinction.  

\begin{figure}[tbh]
{\includegraphics[width=8truecm]{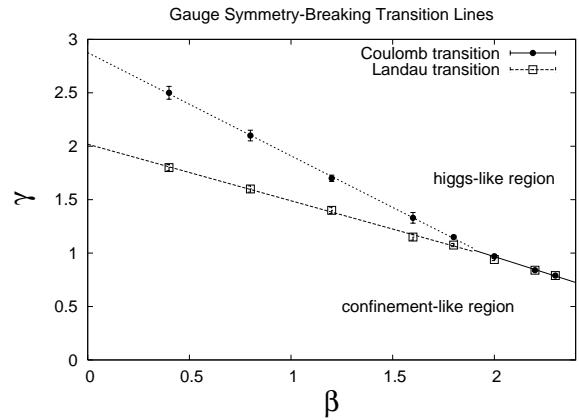}}
\caption{The location of remnant global gauge symmetry breaking in 
Landau and Coulomb gauges, in the $\b-\g$ coupling plane.} 
\label{phase}
\end{figure}

\begin{figure*}[htb]
\subfigure[~]  
{   
 \label{g083}
 \includegraphics[scale=0.6]{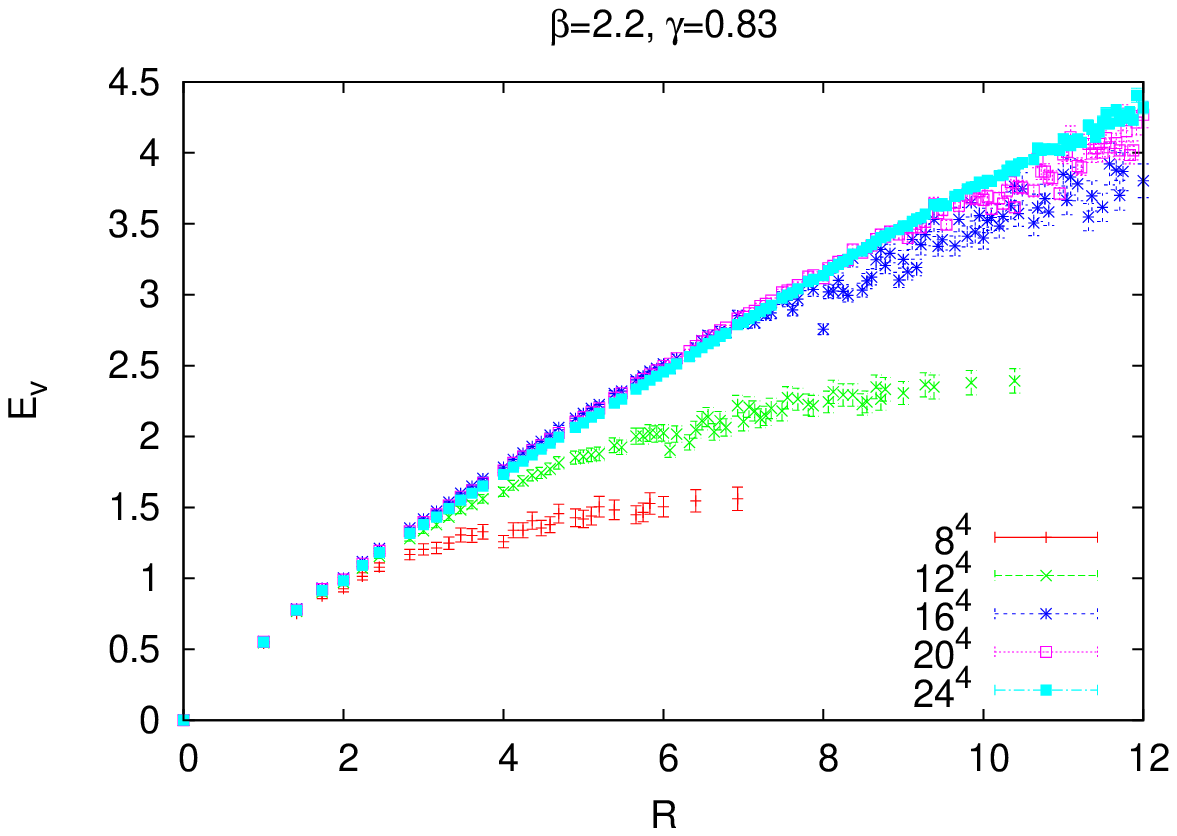}
}
\subfigure[~]
{   
 \label{g084}
 \includegraphics[scale=0.6]{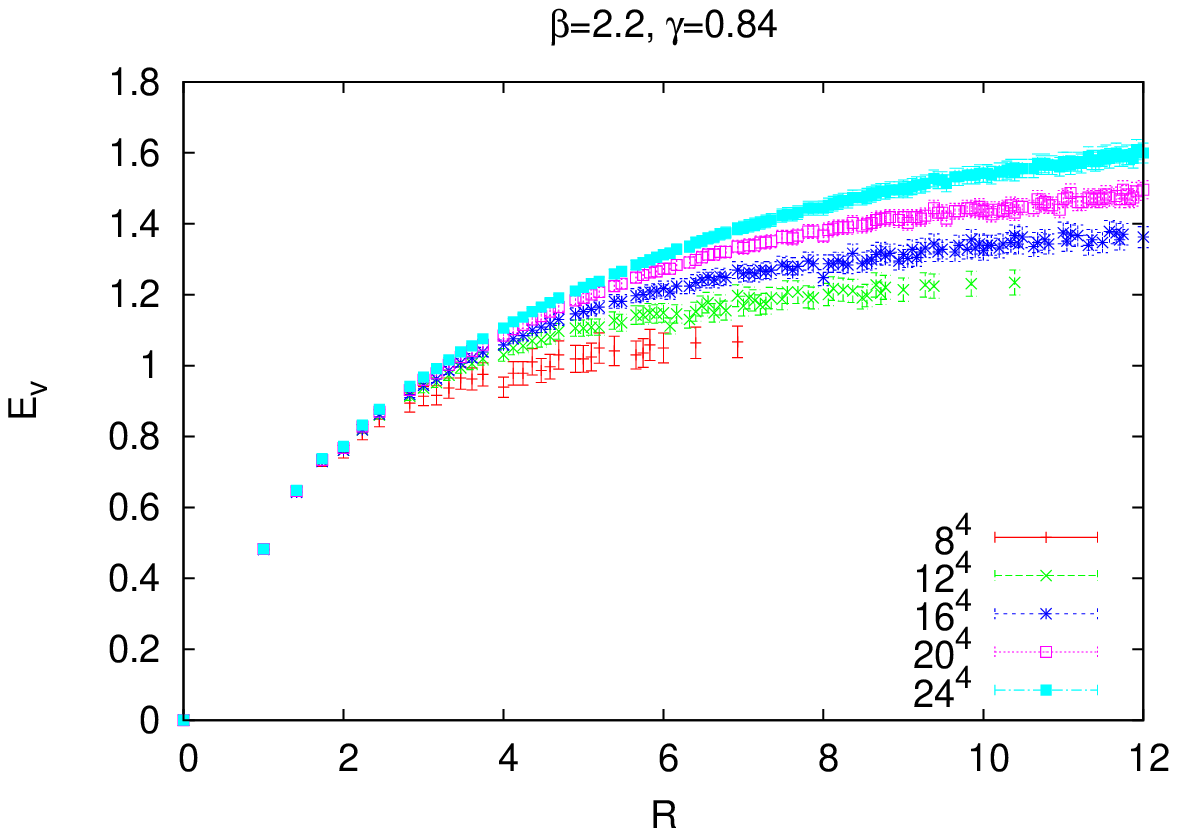}
}
\subfigure[~]
{   
 \label{g085}
 \includegraphics[scale=0.6]{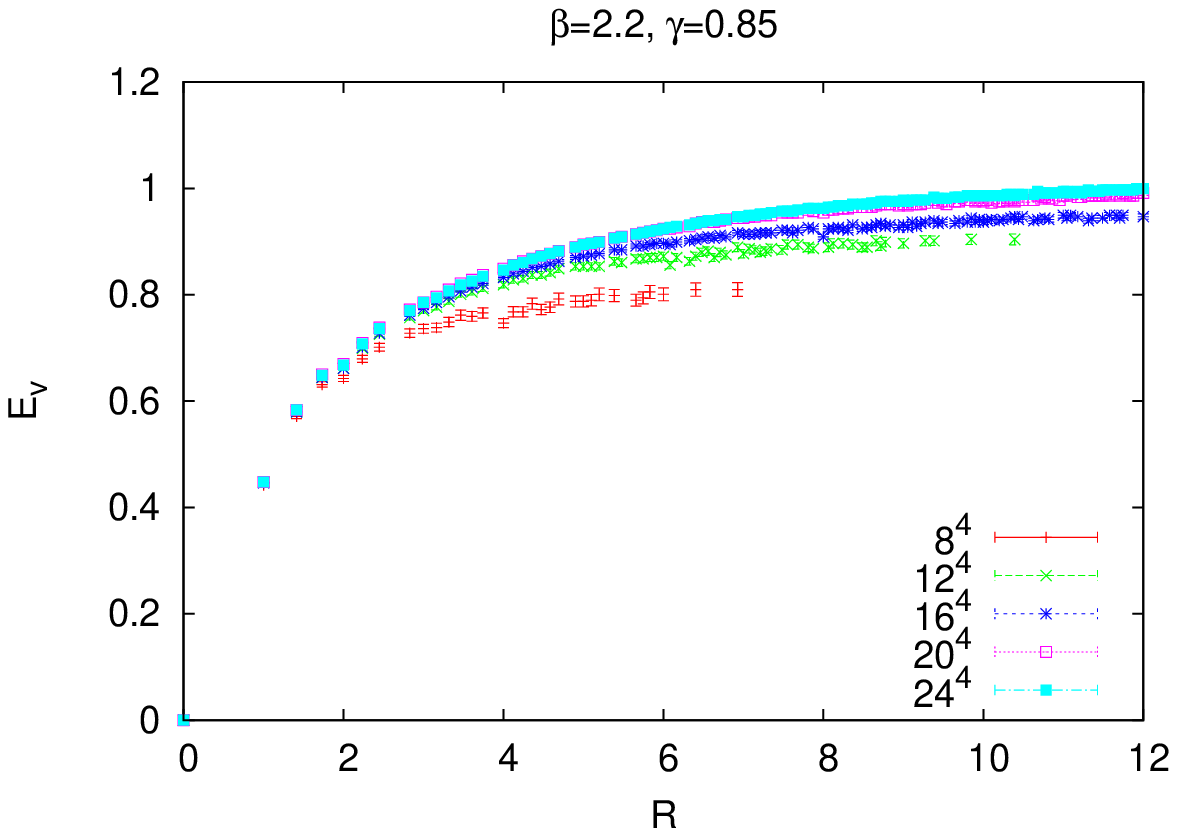}
}
\caption{$E_V(R)$ vs.\ $R$ for the Dirac states in the gauge-Higgs model at $\b=2.2$. (a) just below the remnant symmetry breaking transition at $\g=0.83$; (b) very close to the transition, at $\g=0.84$; (c) just above the transition at $\g=0.85$.} 
\label{b22}
\end{figure*}

    First let us show that S${}_c$-confinement does not exist in part of the phase diagram.  In Fig.\ \ref{phase}  we reproduce a figure
from ref.\ \cite{Caudy:2007sf}.  The dashed and dotted lines will be explained below, but the solid line denotes a line of very rapid crossover behavior which, according to ref.\ \cite{Bonati:2009pf}, turns into a line of 1st order transitions beginning at $\b=2.7266$.  
In Fig.\ \ref{b22} we display the Coulomb energies $E_V(R)$ vs.\ $R$, with $E_V(R)$ defined on the lattice by \rf{Ecoul},  at $\b=2.2$ and $\g=0.83, 0.84,0.85$.    Data is taken at lattice volumes $8^4, 12^4, 16^4, 20^4, 24^4$.  There is clearly a drastic
change in behavior around $\g=0.84$, which marks the transition from the confinement-like to the Higgs region at this $\b$ value.
We see that in the confinement-like region, at $\g=0.83$ just below the transition, the data tends to a straight line as the volume increases.  While not a proof, this is consistent with the hypothesis of S${}_c$-confinement, in that even the quasi-factorizable state 
\rf{Psifact} has an energy which rises linearly with separation, at least at large volumes.  On the other hand, only raising the value of $\g$ slightly to 
$\g=0.85$,  the energy in the Higgs region tends to a constant with separation, and S${}_c$-confinement is lost.  This fact
demonstrates that S${}_c$-confinement cannot exist everywhere in the $\b-\g$ plane.  At $\g=0.84$, which is approximately the transition point,
the data does not seem to have quite converged to its large volume limit.

\begin{figure}[tbh]
 \includegraphics[scale=0.6]{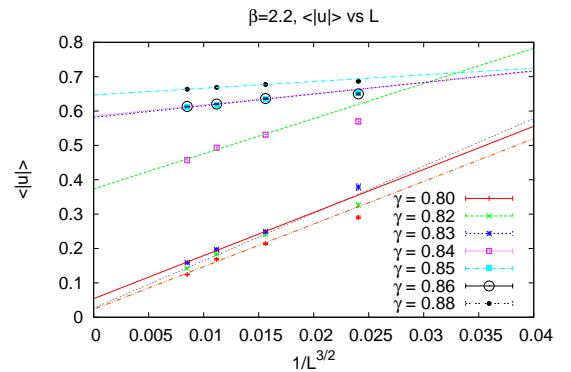}
\caption{The remnant symmetry order parameter $\langle |u| \rangle$ at $\b=2.2$ vs.\ the inverse square root of
3-volume, $L^{-3/2}$.   A rough extrapolation to the infinite volume limit is made by fitting a straight line to the three largest volume
data points.  Note the sudden jump in the extrapolated value between $\g=0.82$ and $\g=0.84$, indicating the breaking
of the remnant gauge symmetry.} 
\label{u22}
\end{figure}

\begin{figure}[tbh]
 \includegraphics[scale=0.6]{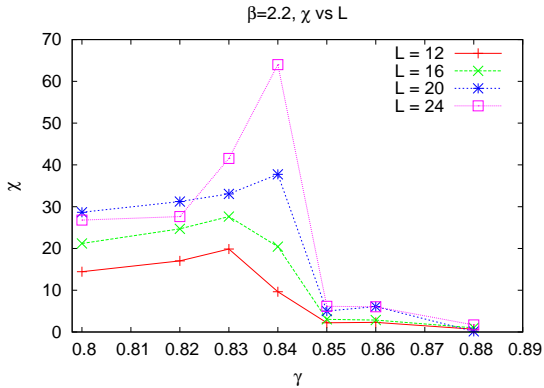}
\caption{Susceptibility $\chi$ vs.\ $\g$ in the $|u|$ order parameter at $\b=2.2$, at various lattice volumes of extension $L$.   The sharp peak at $\g=0.84$ corresponds to the remnant symmetry-breaking transition.} 
\label{x22}
\end{figure}

    The fact that $E_V(R)$ loses S${}_c$-confinement in the Higgs region, for $V(\vx,\vy;A)$ given by \rf{VCoul}, is associated with the breaking of a remnant global symmetry that exists in Coulomb gauge, and in this connection we must now go over some of the same ground covered in ref.\ \cite{Caudy:2007sf}, although this time from a slightly different perspective.  The global symmetry in question is a symmetry under gauge transformations $g(\vx,t)=g(t)$ which are time-dependent but constant in three-space.  Coulomb gauge is preserved under such transformations.  The appropriate order parameter for the symmetry breaking on a time slice is
\beq
              u(t) =  {1\over \sqrt{2} V_3} \sum_{\vx} U_0(\vx,t) \ ,
\label{ut}
\eeq
where $V_3$ is the lattice volume of the time slice, and we define the susceptability (somewhat differently from \cite{Caudy:2007sf}) as
\beq
       \chi = V_3(\langle |u|^2 \rangle - \langle |u| \rangle^2) \ ,
\eeq
where
\beq
          |u| = \sqrt{ {1\over N_t} \sum_{t=1}^{N_t} \tr [u^\dg(t) u(t)]} \ ,
\eeq 
and $N_t$ is the lattice extension in the time direction.
In the S${}_c$-confinement region it is necessary that $\langle |u| \rangle \ra 0$ as $V_3 \ra \infty$.  Then if there are only finite-range correlations,  we expect that the Coulomb energy $E_V(R)$ would be asymptotically linear at large $R$ in the infinite volume limit, which seems indeed to be the case in Fig.\ \ref{g083}.  Conversely, if $\langle |u| \rangle$ tends to a non-zero constant in the infinite volume limit, then it is natural to expect that $E_V(R)$ flattens out and becomes constant as $R\ra \infty$, which is what we see in Fig.\ \ref{g085}.  To confirm this interpretation it is useful to plot $\langle |u|\rangle$ at $\b=2.2$ versus the expected volume dependence $L^{-3/2}$ on a hypercubic lattice of extension $L$.  The data is displayed, for $\g$ values in the 
neighborhood of the transition, in Fig.\ \ref{u22}, along with a straight-line best fit to the three largest-volume data points at each $\g$.  The straight-line fits to the data at $\g<0.83$ extrapolate to values slightly above zero, but this is probably due to our modest lattice sizes.  At 
$\g \ge 0.84$, the extrapolation is clearly to values well above zero.
This looks very much like a first order global symmetry-breaking transition, with the transition point at roughly $\g=0.84$, although of course we cannot rule out a very sharp continuous transition.  In Fig.\ \ref{x22} we display the susceptibility $\chi$ vs.\ $\g$ at various
lattice volumes, and again the growing peak at $\g=0.84$ is consistent with a transition at that point.  As for thermodynamic behavior,
the plaquette energy vs.\ $\g$ has a large slope at the transition \cite{Caudy:2007sf}, although given the results reported in 
\cite{Bonati:2009pf} this is probably due to a sharp crossover rather than a thermodynamic transition.

\begin{figure}[tbh]
 \includegraphics[scale=0.6]{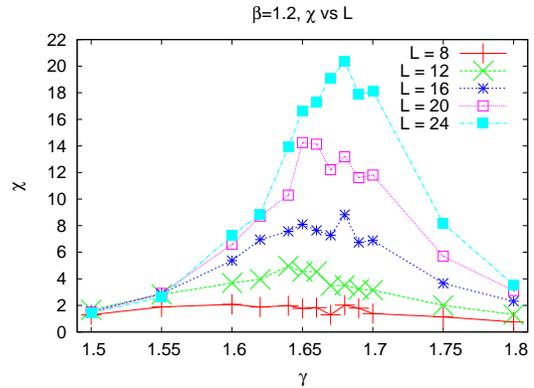}
\caption{Same as Fig.\ \ref{x22}, at $\b=1.2$.  The peak in the susceptibility is much broader (by an order of magnitude) than
the corresponding peak at $\b=2.2$.} 
\label{x12}
\end{figure}

\begin{figure}[tbh]
 \includegraphics[scale=0.6]{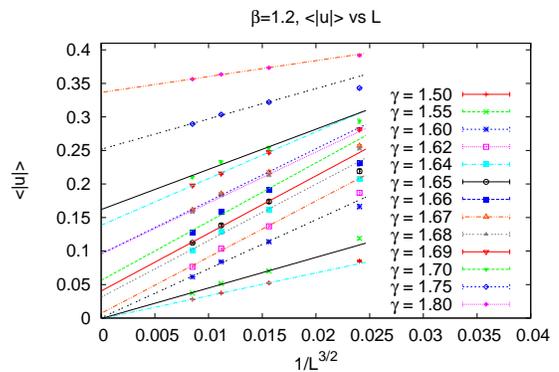}
\caption{Same as Fig.\ \ref{u22} at $\b=1.2$.  Instead of a sharp discontinuity at the transition, the order parameter appears to
increase smoothly away from zero, somewhere in the neighborhood of $\g \approx 1.62$.  This is reminiscent of a continuous
transition, although there is in fact no thermodynamic transition, or even a rapid crossover, at this point.} 
\label{u12}
\end{figure}

\begin{figure*}[htb]
\subfigure[~]  
{   
 \label{g155}
 \includegraphics[scale=0.6]{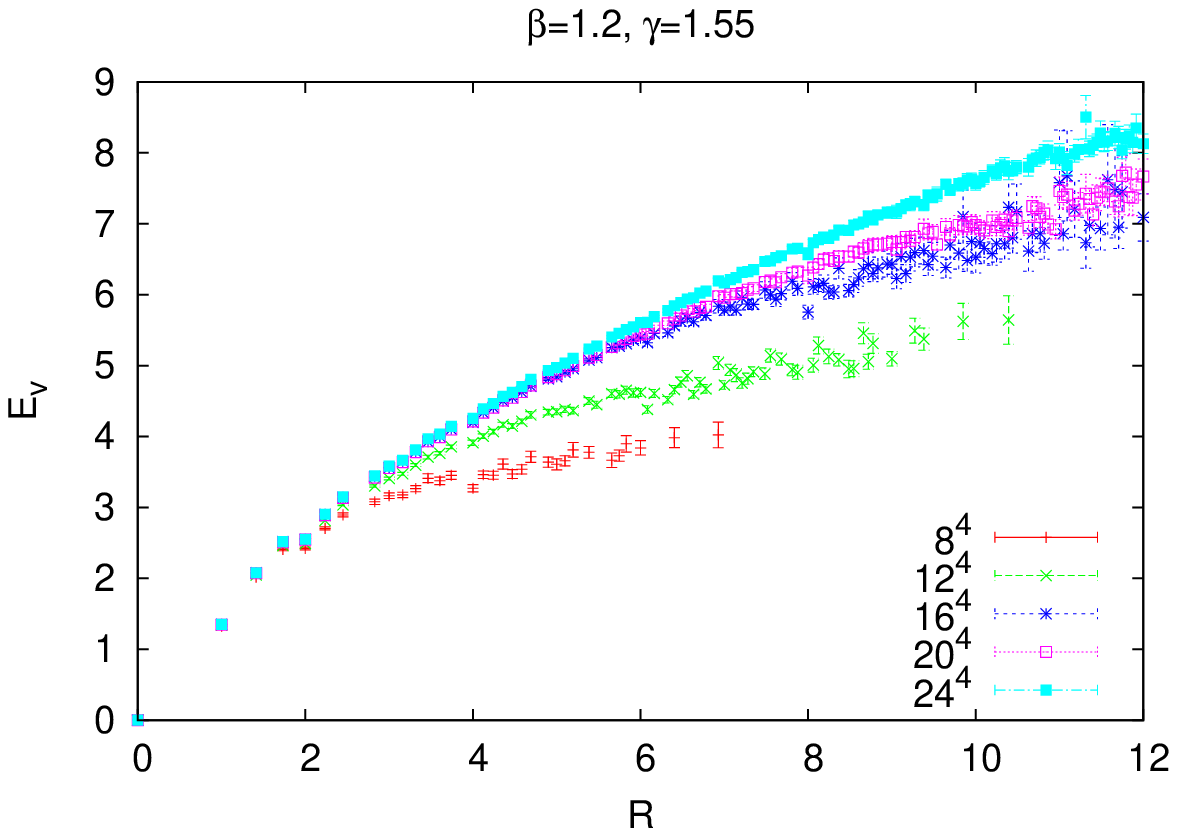}
}
\subfigure[~]
{   
 \label{g168}
 \includegraphics[scale=0.6]{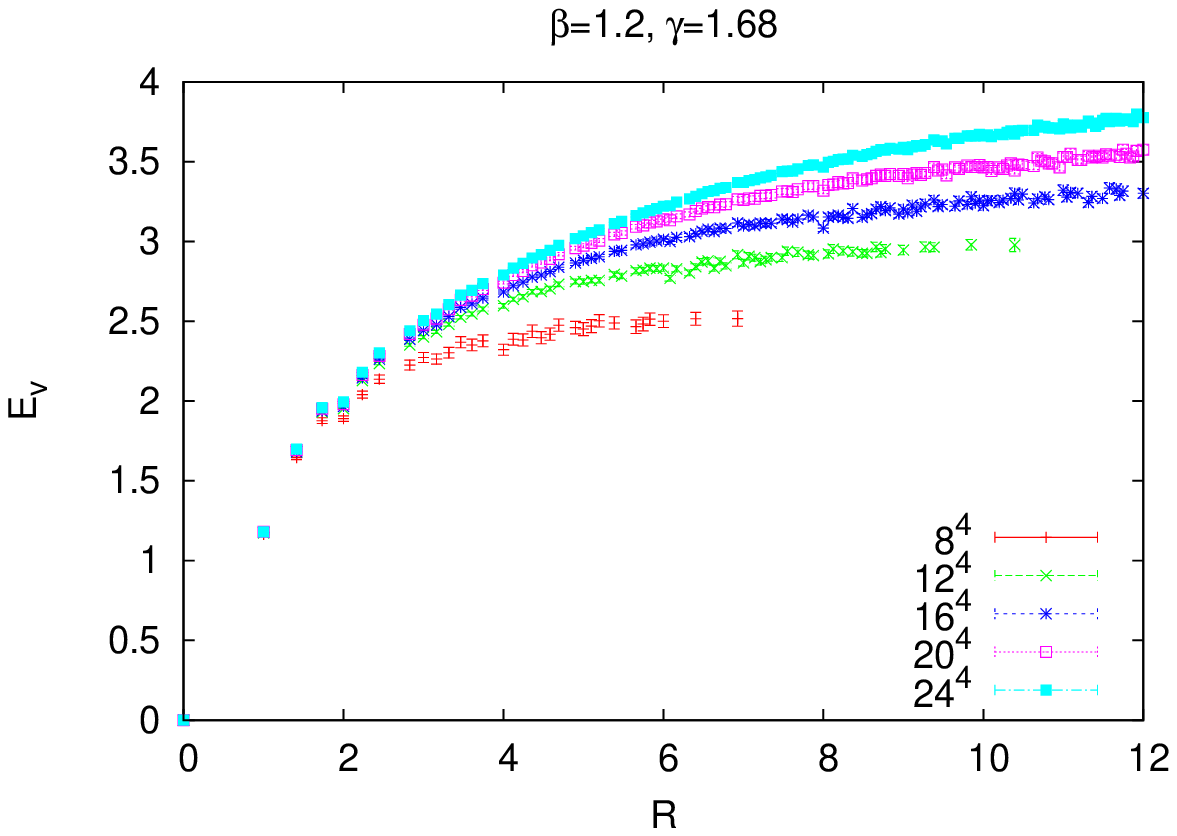}
}
\subfigure[~]
{   
 \label{g1755}
 \includegraphics[scale=0.6]{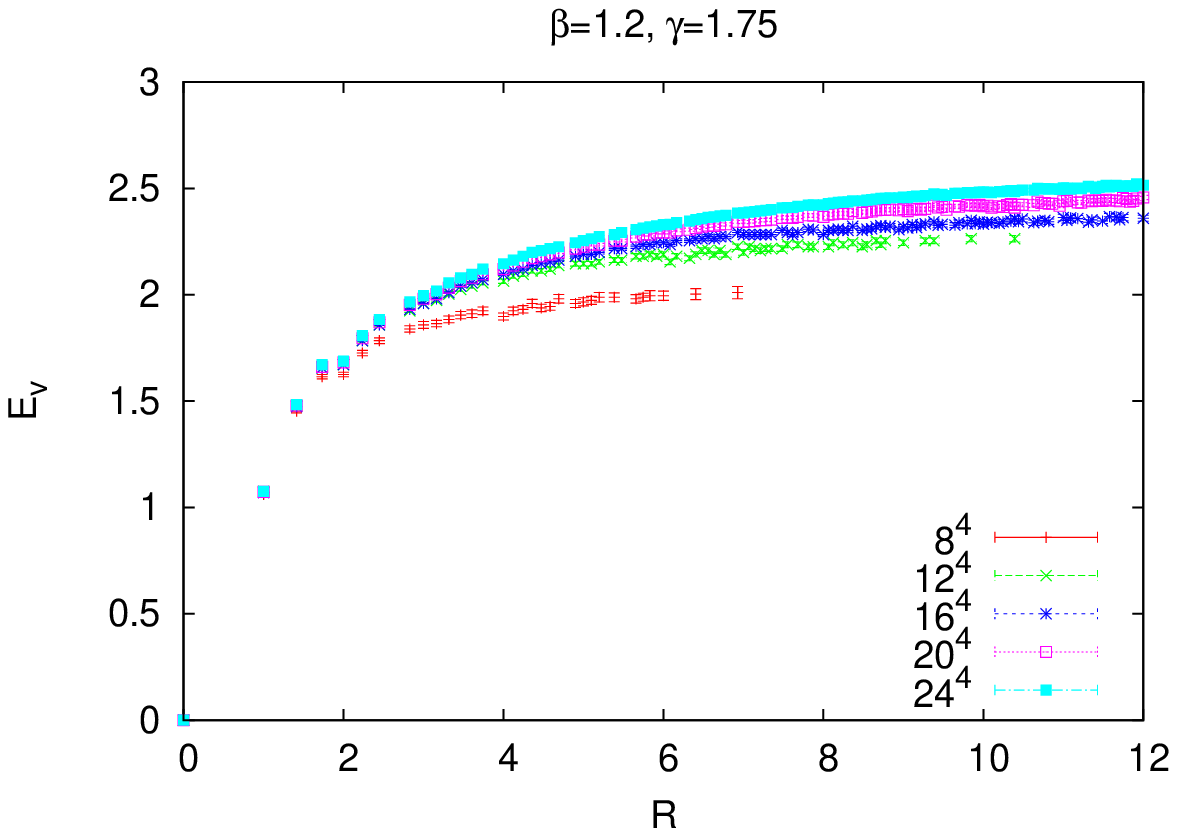}
}
\caption{$E_V(R)$ vs.\ $R$ for the Dirac states in the gauge-Higgs model, as in Fig.\ \ref{b22}, but this time at $\b=1.2$. (a) below the (continuous) remnant symmetry breaking transition at $\g=1.55$; (b) close to, but slightly above the transition, at $\g=1.68$; (c) above the transition at $\g=1.75$.} 
\label{b12}x
\end{figure*}

    In view of the FSOS theorem it is interesting to repeat these calculations at a value of $\b$ in which there is no sign of a thermodynamic transition, or even a sharp crossover.  So we study the behavior of $|u|$ and the susceptability at $\b=1.2$ in the range $1.4\le \g \le 1.8$.
We find a peak in the susceptability at roughly $\g=1.68$ (Fig.\ \ref{x12}), but the peak is broader than the peak at $\b=2.2$ by about an order of magnitude.  The order parameter $\langle |u| \rangle$ in the region of the peak is displayed in Fig.\ \ref{u12}, and here we see that the value of 
$\langle |u| \rangle$, extrapolated to infinite volume, seems to rise smoothly away from zero, without the sudden jump apparent in the
corresponding figure \ref{u22} at $\b=2.2$.  This behavior is consistent (in accord with earlier work \cite{Caudy:2007sf}) with a continuous global symmetry-breaking transition in a region where there is no thermodynamic transition.  This is not a paradox, nor does it violate FSOS.  The reason is that the order parameter $|u|$ is a highly non-local operator (since its definition involves fixing to Coulomb gauge), and operators of that kind can have
non-analytic behavior even if such non-analyticity is absent in the free energy.  
In Figure \ref{b12} we show $E_V(R)$ at $\b=1.2$ just below the symmetry-breaking transition, at the transition, and just above the transition, at $\b=1.55,1.68, 1.75$ respectively.  The behavior is similar to Fig.\ \ref{b22}.

    Of course, other gauges have other remnant symmetries.   In Landau gauge
one has the symmetry under gauge transformations which are constant in spacetime, i.e.\ $g(\vx,t)=g$, and also under certain 
non-constant transformations with a very special dependence on spacetime (cf.\ Hata \cite{Hata:1983cs,*Hata:1981nd} and 
Kugo \cite{Kugo:1995km}).
This remnant symmetry in Lorentz gauge also breaks spontaneously in the gauge-Higgs theory, but the Coulomb gauge remnant symmetry and the Landau gauge remnant symmetry break in different places in the $\b-\g$ plane.  The transition lines for both
gauges are displayed in Fig.\ \ref{phase}, with the two lines joining (and coinciding with a sharp crossover or first-order
transition in thermodynamic quantities) for $\b>2$.   Because the location of the spontaneous breaking of remnant symmetry
is gauge dependent, it was argued in ref.\ \cite{Caudy:2007sf} that such symmetry breaking is physically irrelevant.
The S${}_c$-confinement criterion forces us to adopt a more nuanced view, at least of the spontaneous breaking of remnant symmetry
in Coulomb gauge.  We do not claim that the transition line for this remnant symmetry necessarily denotes the transition from
S${}_c$-confinement to C-confinement.  That may or may not be true, since S${}_c$-confinement in the operator $V(\vx,\vy,A)$ of \rf{VCoul} is a 
necessary but not a sufficient condition for S${}_c$-confinement in the gauge-Higgs system.  What we are able to conclude is that
if S${}_c$-confinement exists in any region of the $\b-\g$ plane, then there is necessarily a transition line {\it somewhere} in the phase diagram which completely isolates the confinement-like region with the S${}_c$-confinement property from the Higgs region with only the C-confinement property.

\subsection{The pseudo-matter transition}

   If the energy $E_V(R)$ of the pseudo-matter state (corresponding to the operator $V(\vx,\vy,A)$ in \rf{pseudo}) rises without limit as $R$ increases, then this implies that the following operator constructed from the pseudo-matter field
\bea
    O_{pm} &\equiv&   {\langle  \vph_1^\dg(\vx,t ) U_0(\vx,t) \vph_1(\vx,t+1) \rangle
                              \over \langle  \vph_1^\dg(\vx,t ) \vph_1(\vx,t)  \rangle }  \non \\
                  &=&  V_3  \langle  \vph_1^\dg(\vx,t ) U_0(\vx,t) \vph_1(\vx,t) \rangle
\label{Opm}
\eea
vanishes, where we have used the fact that $\sum_\vx \vph_1^{a\dg}((\vx,t) \vph^a_1(\vx,t) = 1$.
This means that the gauge-singlet expression $O_{pm}$, like $u(t)$ in eq.\ \rf{ut}, can be used as a (highly non-local) order parameter.
If $O_{pm} \ne 0$, then the system is in the C-confined phase.  The condition $O_{pm}=0$ is a necessary but not a sufficient condition
for S${}_c$-confinement, as already discussed.

\begin{figure}[t!]
\subfigure[~]  
{   
 \label{Opm22}
 \includegraphics[scale=0.6]{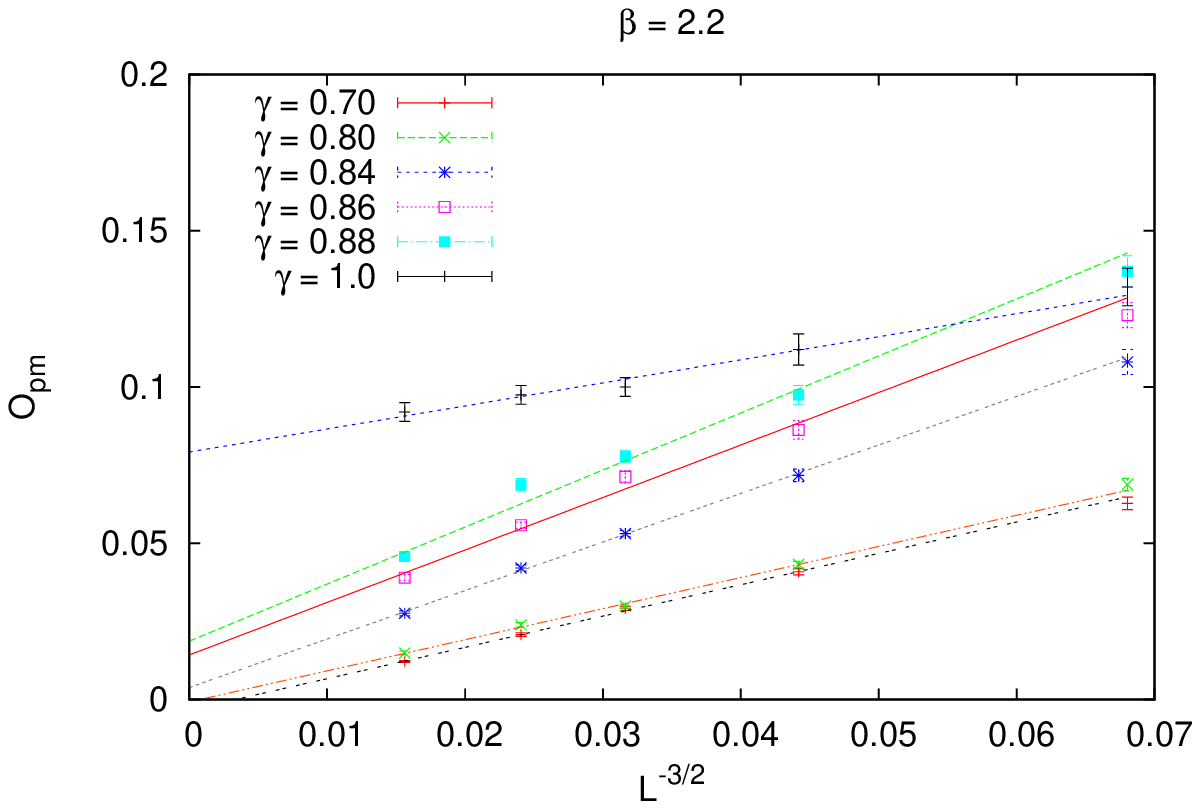}
}
\subfigure[~]
{   
 \label{Opm12}
 \includegraphics[scale=0.6]{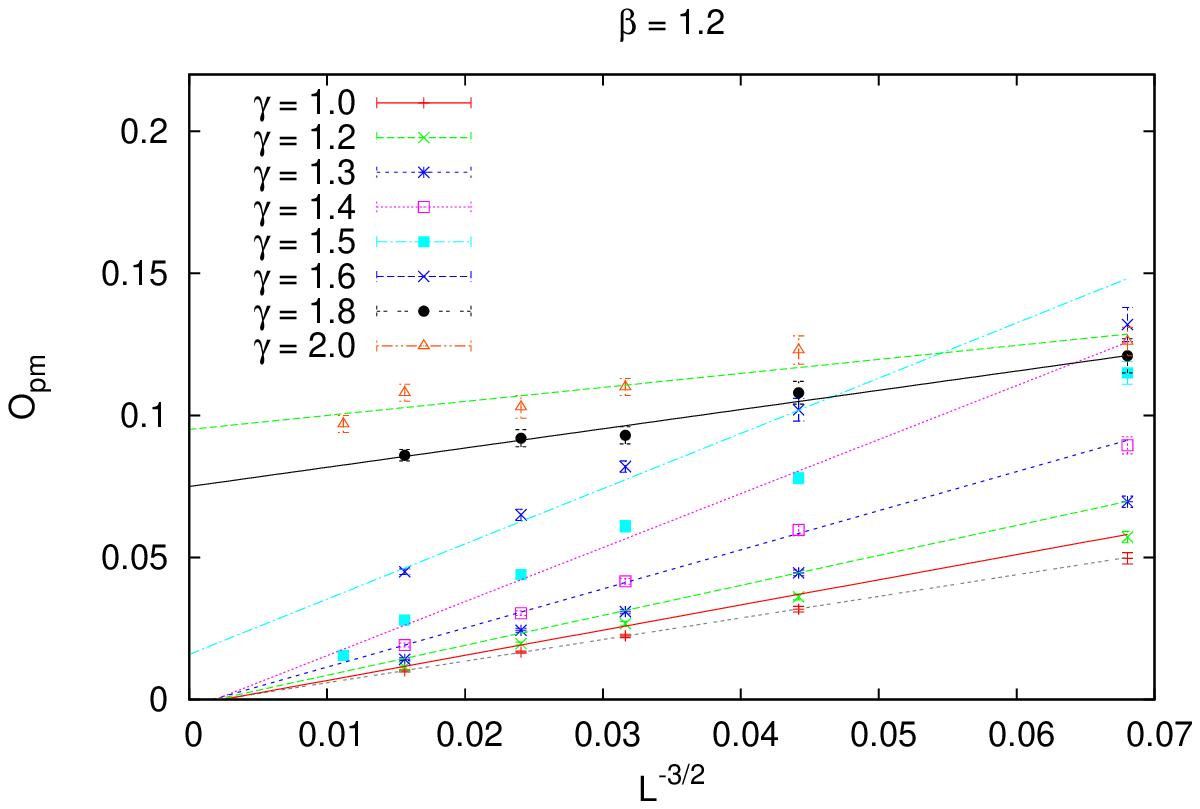}
}
\caption{The pseudo-matter order parameter $O^L_{pm}$, defined in \rf{OpmL} at (a) $\b=2.2$, and (b) $\b=1.2$, plotted vs.\ the inverse
square root of the spatial volume $L^{-3/2}$, at a set of $\g$ values. In both cases the order parameter reliably extrapolates
to a positive value only at the higher $\g$ values.} 
\label{Opm_plots}
\end{figure}

   At finite volume $V_3=L^3$ we define
\bea
    O^L_{pm} &=& V_3 \left\langle {1\over N_t} \sum_{t=1}^{N_t} {1\over V_3} \left| \sum_x \vph_1^\dg(\vx,t ) U_0(\vx,t) \vph_1(\vx,t)
                          \right| \right\rangle  \non \\
                  &=& O_{pm} + O(L^{-3/2}) \ ,
\label{OpmL}
\eea
where $O_{pm}$ is the infinite volume limit.  The results at $L=6,8,10,12,16$  are shown for $\b=2.2$ in Fig.\ \ref{Opm22}.   At $\g=0.8$ and below, the results for $|O^L_{pm}|$ extrapolated to infinite volume are consistent with $O_{pm}=0$, while at $\g=0.9$ and above, the extrapolated value is non-zero.  
Unlike $\langle |u| \rangle$ at $\b=2.2$ in Fig.\ \ref{u22}, the increase in $O_{pm}$ away from zero is gradual as $\g$ increases, in fact it is more like the behavior seen for $\langle |u| \rangle$ at $\b=1.2$ in Fig.\ \ref{u12}.  The same plot, for $\b=1.2$, is shown in Fig.\ \ref{Opm12}, which shows similar behavior.\footnote{Actually at small $\g$ the extrapolation to infinite volume is slightly negative.  But this cannot be the true infinite volume limit, since by definition $O^L_{pm}$ is strictly positive at all $L$, so we blame the slight negative deviation from zero on our relatively small lattice volumes.  In any case the negative deviation is statistically not very significant, e.g.\ at $\g=1.5$ we find an extrapolated value $O_{pm}=-0.004 \pm 0.003$.}   

\begin{figure}[htb]
 \includegraphics[scale=0.6]{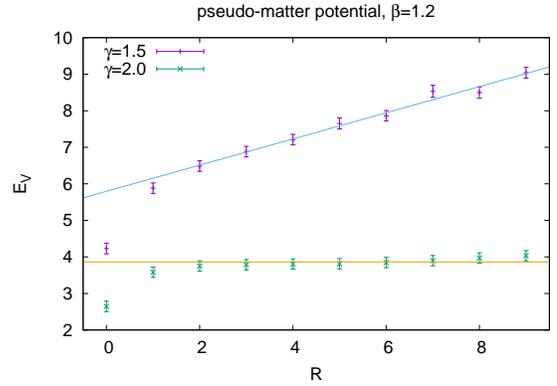}
\caption{The energy $E_V(R)$ defined in \rf{Epm} vs.\ $R$ for pseudo-matter states in the gauge-Higgs model at $\b=1.2$, compared at 
$\g=1.5$ (confining region) and $\g=2.0$ (Higgs region), and computed on a $20^4$ lattice volume. The straight lines are best fits of the data at $R \ge 2$ to a linear rise ($\g=1.5$) and a constant ($\g=2.0$). } 
\label{clap}
\end{figure}

   The energy $E_V(R)$ at $\b=1.2$ on a $20^4$ lattice volume is plotted in Fig.\ \ref{clap} below ($\g=1.5$) and above ($\g=2.0$) the transition point where $O_{pm}$, extrapolated to infinite volume, becomes positive.  The data seems to follow a linear rise below the infinite volume transition point, consistent with S${}_c$-confinement in this region, and is flat, or very nearly so, above the transition, consistent with C-confining behavior.

   It is interesting to contrast Fig.\ \ref{Opm12} with the corresponding figure for the $u$ order parameter in Fig.\ \ref{u12}.  From these figures it seems that the two order parameters, at $\b=1.2$, may change their behavior (i.e.\  transition away from zero) at slightly different values of $\g$.  For the remnant symmetry breaking order parameter, it would appear from the extrapolations in Fig.\ \ref{u12} that the transition is somewhere in the range 
$1.62 < \g < 1.64$, while for $O_{pm}$ the transition seems to be in the range $1.5 < \g < 1.6$.  Of course the small discrepancy may
be due to some small error in the infinite volume extrapolation, but we  should stress that there
is no reason that the transition from S-to-C confining behavior has to occur at exactly the same place in the phase diagram for every operator $V$.  S${}_c$-confinement in the
gauge-Higgs theory means that there exists some limiting boundary in the phase diagram, not identical with $\g=0$, 
such there is no choice of $V$ which has C but not S${}_c$-confining behavior for points in the $\b-\g$ plane below that boundary, and in fact it is interesting that operators as different as $u$ and $O_{pm}$ show a transition at very nearly the same $\g$ value, since there is no thermodynamic transition, or even crossover behavior, in this region.   

\subsection{Wilson line operators}

    Not every choice of $V(\vx,\vy,A)$ is sensitive to the S${}_c$- to C-confinement transition.  We have already argued that if  
$V(\vx,\vy,A)$ is taken to be a simple
Wilson line operator running between points $\vx$ and $\vy$, then this will lead to a quark-antiquark state whose energy will
grow linearly with $R$, even in an abelian theory.  But it is worth checking this assertion in the gauge-Higgs theory.
For lattice sites with an on-axis separation $\vy = \vx + R\hat{k}$,  we define on a timeslice
\beq
       V_{thin}(x,y;A) = U_k(x)U_k(x+\hat{k})...U_k(x+(R-1)\hat{k}) \ ,
\eeq
and 
\beq
          \Psi_{thin}(R) = \overline{q}(x) V_{thin}(x,y;A) q(y) \ ,
\label{Pthin}
\eeq    
with energy expectation value above the vacuum energy
\bea
\lefteqn{E_{thin}(R)} & & \non \\
& & = -\log \left[ \oh \tr[ V_{thin}(x,y,A) U_0(y) V_{thin}^\dg(x+\hat{t},y+\hat{t};A) U^\dg_0(x)] \right]  \ . \non \\
\eea

    It is also interesting to go a little further, and consider an operator which has been used for noise reduction in various studies of lattice gauge theory, see e.g.\ \cite{Bali:1994de}.  This is a Wilson line operator built on the lattice from ``fat'' links, and is it constructed by an iterative procedure.  Begin by defining the initial set of link operators 
$U^{(0)}_k(x)=U_k(x)$ to be the spatial links of the lattice gauge theory.  Then the ``fat'' link operators in SU(2) lattice gauge theory, 
at the $n+1$-th iteration (or ``smearing step''), are constructed from the fat links at the $n$-th smearing step according to the rule
\bea
\lefteqn{U^{(n+1)}_i(x)} & & \non \\
 &=& \N \Bigl( \a U^{(n)}_i(x) + \sum_{j\ne i} \left( U^{(n)}_j(x) U^{(n)}_i(x+\hat{j}) U^\dg_j(x+\hat{i})   \right.  \non \\
 & & \qquad \left.  + U^{(n)\dg}_j(x-\hat{j}) U^{(n)}_i(x-\hat{j}) U^{(n)}_j(x-\hat{j}+\hat{i})  \Bigr) \right) \ ,
\eea           
where $\N$ is chosen to preserve the SU(2) condition $\det[U^{(n+1)}_i(x)]=1$, and the sum is over spatial directions $j=1,2,3$.  
Denote the link operators after the final iteration as $U^{fat}_i(x)$.  For $\vy = \vx + R\hat{k}$ we define on a timeslice
\beq
       V_{fat}(x,y;A) = U^{fat}_k(x)U^{fat}_k(x+\hat{k})...U^{fat}_k(x+(R-1)\hat{k}) \ .
\label{Vfat}
\eeq
In terms of the original link variables, the operator $V_{fat}$ boils down to a superposition of Wilson lines of the general form \rf{super}, and after a large number of smearing steps it is a non-local operator.  The energy expectation value $E_{fat}(R)$ of the state
\beq
          \Psi_{fat}(R) = \overline{q}(x) V_{fat}(x,y;A) q(y)
\label{Pfat}
\eeq
above the vacuum energy is given by
\bea
\lefteqn{E_{fat}(R)} & & \non \\
& & = -\log \left[ \oh \tr[ V_{fat}(x,y,A) U_0(y) V_{fat}^\dg(x+\hat{t},y+\hat{t};A) U^\dg_0(x)] \right] \ .  \non \\
\eea

\begin{figure*}[t!]
\subfigure[~confinement phase]  
{   
 \label{fat83}
 \includegraphics[scale=0.6]{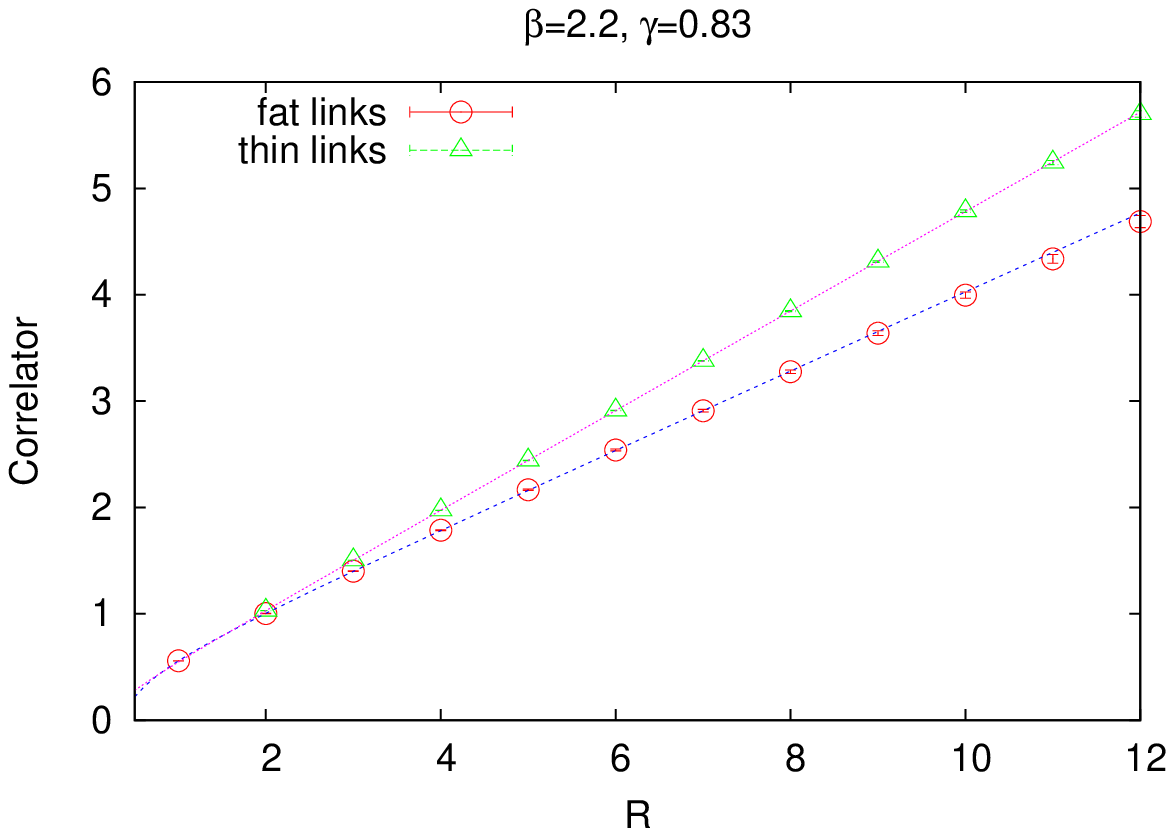}
}
\subfigure[~Higgs phase]
{   
 \label{fat85}
 \includegraphics[scale=0.6]{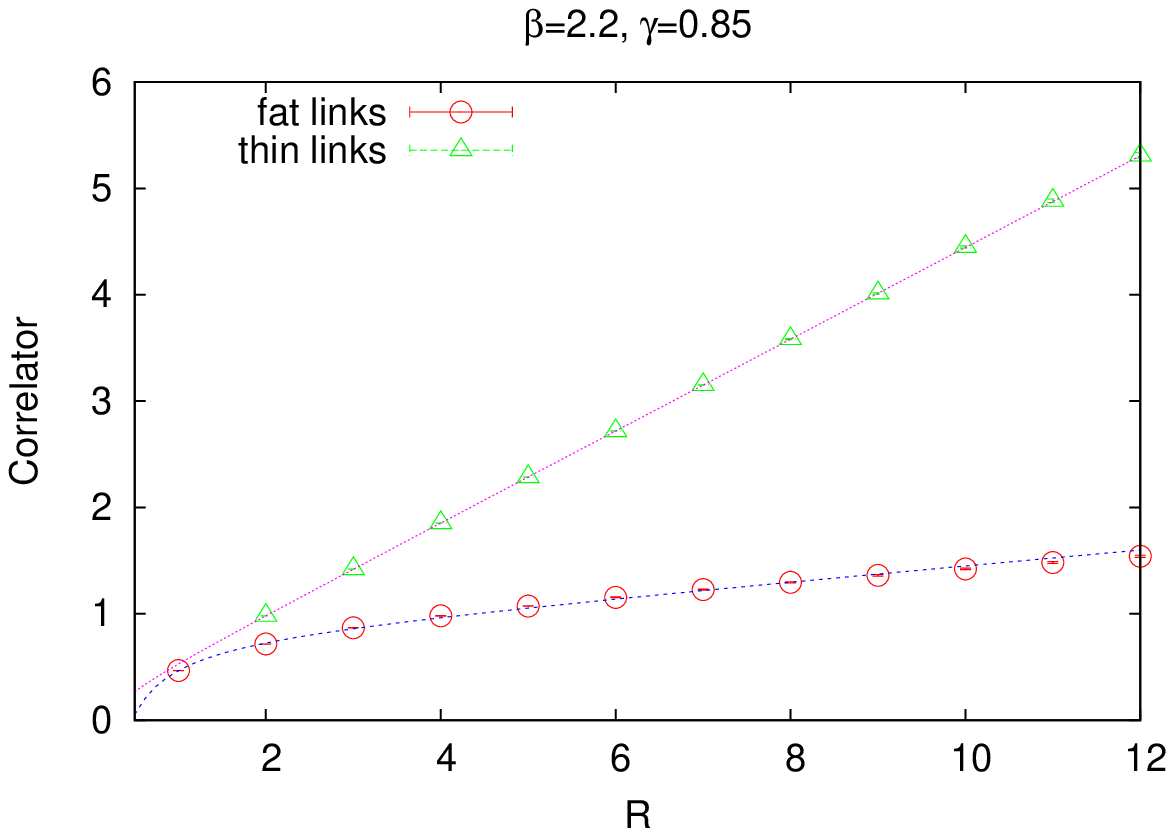}
}
\caption{$E_V(R)$ vs.\ $R$ for states created by Wilson line operators in the gauge-Higgs model at $\b=2.2$.  The figure compares
Wilson line operators obtained from unmodified (``thin'') link variables and ``fat'' link variables.  
(a) just below the crossover transition in the ``confinement phase,'' at $\g=0.83$; (b) just above the transition at $\g=0.85$. These operators,
in contrast to the Dirac and pseudo-matter operators, do not detect the loss of S${}_c$-confinement above the transition, since even in the
Higgs region the energies rise linearly.  But S${}_c$-confinement requires that {\it all} $V$ operators, not just some operators, satisfy the S${}_c$-confinement condition, as explained in the text, otherwise the
system is in the C-confinement phase.}
\label{fatlink}
\end{figure*}

   For testing the fat link operator we follow the old work of Bali et al.\ \cite{Bali:1994de} and construct fat links using $\a=2$ with 100 smearing steps. We might suspect that such an operator is sufficiently non-local to evade the FSOS theorem, and show an S${}_c$- to C-confinement transition. However, this idea is not supported by the numerical results.  In Fig.\  \ref{fat83} we show $E_{thin}(R)$ and $E_{fat}(R)$ at
$\b=2.2, \g=0.83$, which is just below the sharp crossover at $\b=2.2,\g=0.84$ in the S${}_c$-confinement phase.  Both the the thin and fat link
Wilson line operators result in a linear increase in energy with separation. The data in Fig.\ \ref{fat85} is taken just above the crossover,
at  $\b=2.2,\g=0.85$, in the C-confining
Higgs phase.  While the thin-link Wilson line operator shows little change across the transition, there is quite a large change in energy associated with the fat-link Wilson line operator.  Nevertheless, although the slope of $E_{fat}$ vs.\ $R$ has been reduced by about a factor
of five, $E_{fat}(R)$ still seems to rise roughly linearly with $R$, and, like $E_{thin}(R)$ associated with a simple Wilson line, cannot be
used to detect the transition of S${}_c$- to C-confinement.  Again, this is not a paradox; it is easy to construct physical states whose energy
expectation value rises linearly with charge separation, even in non-confining theories such as QED, and both the thick and thin Wilson lines generate states of that type in the gauge-Higgs theory.  In a C-confining phase there must
however be {\it some} physical states whose energy does not rise in this way, and we have seen that the Dirac and pseudo-matter states
in the Higgs phase are examples with this property.
   
\section{Remarks on other criteria}

    Other criteria for confinement in gauge + matter systems have been suggested in the literature.  This may be the appropriate
place to comment on some of them. 

\subsection{The Kugo-Ojima criterion}

    The Kugo-Ojima criterion is essentially a condition which guarantees C-confinement.  The problem is that this condition
is violated at the non-perturbative level in the gauge-Higgs model we are considering, yet C-confinement persists throughout
the phase diagram.  This calls into question the utility of the Kugo-Ojima criterion, at least in the lattice model under 
consideration.
 
    Kugo and Ojima \cite{Kugo:1979gm} introduce the function $u^{ab}(p^2)$, defined by the expression
\bea
 \lefteqn{u^{ab}(p^2) \left( g_{\m\n} - {p_\m p_\n \over p^2} \right) = }
 \non \\
           & &   \int d^4 x ~ e^{ip(x-y)} \langle 0|T[D_\m c^a(x) 
           g(A_\n \times \overline{c})^b(y) | 0 \rangle \ ,
\eea
where $c,\overline{c}$ are the ghost-antighost fields.
They then show that the expectation value of color charge in any physical
state vanishes  
\beq
           \langle \mbox{phys}\ |Q^a |\mbox{phys}\rangle = 0 \ ,
\label{Qa}
\eeq 
providing that (i) remnant symmetry with respect to spacetime-independent
gauge transformations is unbroken; and (ii) the following condition is satisfied: 
\beq
           u^{ab}(0) = -\d^{ab} \ .
\label{kg}
\eeq
This latter condition is the Kugo-Ojima confinement criterion.  In fact conditions (i) and (ii) are not
entirely independent.  It was shown by 
Hata in ref.\ \cite{Hata:1983cs,*Hata:1981nd}  (see also Kugo in ref.\ \cite{Kugo:1995km}) that the condition \rf{kg} is a necessary 
(and probably sufficient) condition for the unbroken realization of the residual spacetime-dependent 
symmetry, while an unbroken, spacetime independent
symmetry is required, in {\it addition} to \rf{kg}, for the vanishing of 
$\langle \psi |Q^a |\psi \rangle$ in physical states.   Both of these symmetries 
are necessarily broken if a Higgs field acquires a VEV in Landau gauge.

    The problem here is that in Landau gauge the Higgs field does acquire a VEV above the Landau transition line
shown in Fig.\ \ref{phase}, yet physical states are still color-singlets, as shown by FSOS.  Something then
seems amiss with the Kugo-Ojima condition, as it does not appear to be a necessary condition for color
confinement.

     One possibility is that BRST symmetry, which underlies the Kugo-Ojima approach, does not exist beyond perturbation
theory.  We are referring here to the Neuberger ``0/0'' problem.  The difficulty
pointed out by Neuberger \cite{Neuberger:1986xz} is that the functional integral in a covariant gauge
\beq
            Z = \int DA_\m Dc D\overline{c} \exp[-(S+S_{gf})] \ ,
\eeq
where $S_{gf}$ is the gauge-fixing part of the action, simply vanishes.  The expectation value of any quantity is therefore ill-defined, this is the ``0/0'' problem.  The origin of the problem is the sum over Gribov copies.  These contribute to the
functional integral with both positive and negative signs, and the sum over all copies vanishes.  The obvious resolution,
which is what is actually done for Landau gauge in lattice Monte Carlo simulations, is to restrict the sum over copies to those
for which the Faddeev-Popov determinant is positive.  However, this restriction has the unpleasant feature of breaking BRST 
invariance explicitly, which may account for the failure of the Kugo-Ojima criterion, at least as applied to lattice Monte Carlo 
simulations.  There have been some efforts to address the 0/0 problem in a way which preserves BRST symmetry (see, e.g.,
\cite{vonSmekal:2013cla}), but we
are not aware of a complete resolution of the problem in the case of interest.

\subsection{Unphysical poles and non-positivity in quark/gluon propagators}

     It has been suggested that confinement can be attributed to unphysical poles and/or a loss of reflection positivity in the quark and gluon propagators.  This view is common in some of the Schwinger-Dyson literature;
an early reference is \cite{Stingl:1985hx}.  The idea is that positivity violation or unphysical poles would mean that these color-charged particles could never appear in the initial or final states of scattering amplitudes.  A standard example is that $W$ particles have physical poles in their propagators, while the gluon propagator has unphysical poles, and this would explain why $W$'s are observed while quarks and gluons are not.  This argument is not entirely convincing, in our opinion.  We agree that Landau gauge quark and gluon operators, operating on the vacuum, do not create physical states, and there are a priori reasons for this.  We disagree that this fact implies that widely separated, color-charged particles cannot show up in physical asymptotic states; our interpretation of unphysical poles and non-positivity is simply that quark and gluon operators in covariant gauges are the wrong operators to use in constructing physical states with separated color charges.   
As for the comparison of $W$-particles and gluons, this example may be a little misleading, since the $W$-particles which appear in asymptotic states are, as explained previously, actually color neutral objects created by gauge-invariant composite operators, and the perturbative results in this case can be justified in a gauge-invariant formalism, cf.\ \cite{Frohlich:1981yi,*Frohlich:1980gj} and \cite{Egger:2017tkd,*Torek:2016ede}. 
 
    Regarding physical states and positivity:  Let $G_L(x;A)$ be the gauge transformation which takes an arbitrary gauge field into Landau gauge.  What is actually done in a lattice Monte Carlo calculation of the Landau gauge gluon propagator is to compute the Fourier transform of the gauge-invariant observable
\beq
       D^{ab}_{\m \n}(x-y) = \langle [G_L\circ A]^a_\m(x) [G_L \circ A]^b_\n(y) \rangle \ ,
\eeq
where this time $ab$ are color indices in the adjoint representation, and the gauge transformation $G_L$ always takes $A$ to a gauge
copy within the first Gribov horizon.  But unlike $G_C(\vx;A)$ in Coulomb gauge, the Landau gauge transformation $G_L(x;A)$ depends on the $A$-field throughout the spacetime volume, rather than the $A$-field only on a particular time slice.  This violates the usual requirement for proving the reflection positivity of a correlator, so there seems to be no obvious reason to believe that the Landau gauge gluon propagator should satisfy positivity \cite{Zwanziger:1991gz}, and lattice Monte Carlo simulations show that this property is indeed violated \cite{Cucchieri:2004mf,Bowman:2007du}.  Lack of positivity, from this point of view, should not come as a surprise.

    On the other hand, one may argue that the time non-locality is fictitious, because it can be eliminated by introducing ghost 
fields.  According to this argument the gluon propagator can be expressed as
\beq
          D^{ab}_{\m \n}(x-y) = {1\over Z} \int DA_\m Dc D\overline{c} ~ A^a_\m(x) A^b_\n(y) \exp[-(S+S_{gf})] \ ,
\eeq
and since $S+S_{gf}$ is local in time, the usual reflection positivity argument should apply.  But this construction runs right into
the Neuberger 0/0 problem.  Of course, as in the Monte Carlo simulation, one could demand that the range of the functional 
integral be restricted somehow, e.g.\ to stay within the first Gribov horizon.  As already mentioned, this means that the BRST transformation is no longer a symmetry of the Landau gauge-fixed theory.\footnote{BRST symmetry is also broken in the Gribov-Zwanziger (GZ) action, although in this case it is conjectured that the problem is not fatal to the identification of physical states.
See \cite{Schaden:2014bea} for a discussion of this point, and \cite{Cucchieri:2016bvz} for a lattice investigation of BRST symmetry breaking in connection with the GZ action.}

    If one cannot rely on BRST arguments, then there is no strong reason to suppose that quark and gluon operators in covariant gauges create physical states, and there is one good reason to think otherwise, namely, the fact that correlators of such operators do not satisfy the requirements for reflection positivity.   It is true, of course, that the transverse A-field creates a physical state in the abelian zero coupling limit, where there are no Gribov copies, the 0/0 argument does not apply, and BRST symmetry is unbroken.  But if BRST symmetry is broken
at any non-zero gauge coupling, no matter how small, then a smooth $g\ra 0$ limit to in/out states created by the transverse A-field does not exist.  Similar considerations apply to quark propagators in Landau gauge.

    One should not conclude, however, from the fact that isolated quark/gluon operators in Landau gauge do not create physical states, that physical states with widely separated color charges do not exist.  The correct conclusion is that other types of operators must be employed to create such states.   The whole point of the present article is that there exist color neutral gauge-singlet states containing widely separated color charges, e.g.\ massive quarks and antiquarks, unscreened by matter fields.  The issue is not whether such physical states exist; the Dirac state, the pseudo-matter state, and the fat link state are explicit examples of physical states of that kind.  The real question is whether or not a wide separation of color charged objects in a physical state incurs a proportionally large cost in energy, and this question, which takes us right back to 
S${}_c$-confinement, is not settled by the behavior of quark and gluon propagators in covariant gauges.
 
\subsection{The Fredenhagen-Marcu criterion}

    Consider a contour $C_{\vx \vy}$ between two points $\vx ,\vy$ at equal times, which has the shape of a staple, and $U(C_{\vx \vy})$ is
the Wilson line running along this contour.  The staple has an 
extension $R/2$ in the time direction, and $R=|\vx-\vy|$ in the space direction.   Fredenhagen and Marcu \cite{Fredenhagen:1985ft}
introduce an order parameter
\bea
      \rho &=& \lim_{R\ra \infty} { \Bigl| \sum_i \langle \phi^\dg_i(x)
       U(C_{xy}) \phi_i(y) \rangle \Bigr|^2 \over W(R,R)} \ ,
\label{FM}
\eea
where the sum runs over all matter fields in the theory.  The Fredenhagen-Marcu (FM) confinement criterion is $\r>0$.
The idea is that if the perimeter-law falloff of a Wilson loop is
dominated by charge screening due to matter fields,
then the numerator and denominator
in eq.\ \rf{FM} are comparable, and the ratio $\rho$ is non-zero in the
large $R$ limit.  On the other
hand, if the perimeter-law falloff of the Wilson loop is independent
of charge screening, then the denominator may be
much larger than the numerator,
and $\r \ra 0$ in the limit.

    But like the existence of a mass gap (which is also sometimes regarded as a confinement criterion), the FM criterion does not distinguish between the confinement-like and Higgs regions of a gauge-Higgs theory. Even in the case of a gauge-Higgs theory with the scalar fields in the adjoint representation, where the Higgs and confinement regions are entirely separated by a center symmetry-breaking transition line and where the confinement phase has a Wilson loop area law falloff while the Higgs phase does not, the Higgs phase is still ``confining'' according to the Fredenhagen-Marcu criterion in both the Higgs and the confinement phases  \cite{Azcoiti:1987ua}.  In contrast, the 
criterion introduced in this article does make a qualitative distinction, so far as confinement is concerned, between an S${}_c$-confining phase and a Higgs phase in the gauge-adjoint Higgs theory, for reasons we now discuss.

\subsubsection{S${}_c$-confinement in the gauge-adjoint Higgs theory}

   Let us consider an SU(2) lattice gauge theory with the unimodular Higgs field $|\phi|=1$ in the adjoint representation of the gauge group, with action
\bea
     S &=& \b \sum_{plaq} \oh \mbox{Tr}[UUU^\dg U^\dg]  \non \\
         & & + {\gamma\over 4} \sum_{x,\m} \phi^a(x) \phi^b(x+\hat{\m}) \mbox{Tr}[\sigma^a U_\m(x) \sigma^b U^\dg_\m(x)] \ , \non \\
\label{ahiggs}
\eea
where the $\s^a$ are the Pauli matrices, $a=1,2,3$.  Unlike the gauge-Higgs theory with the Higgs field in the fundamental representation, the action of the gauge-adjoint Higgs theory has a global $Z_2$ center symmetry.  In the phase of unbroken center symmetry, Wilson loops fall off asymptotically with an area law, which implies S${}_c$-confinement for the following reason:  We follow the steps of equations 
[\ref{Wline}-\ref{bicovariant}], with the difference being that the $V_t$ operator in the minimal energy state, obtained by propagation
in Euclidean time $t \ra \infty$, will in general depend on the adjoint Higgs field as well as the gauge field.  But because this is the minimal energy state, with energy $E_0(R)$ which increases linearly as $R \ra \infty$,  if follows that any {\it other} $V$ operator, in particular an
operator $V=V(\vx,\vy,A)$ that depends only on the gauge field, will have an energy greater than $E_0(R)$, and is therefore bounded from
below by a linear potential.  So the center-symmetric phase of a gauge adjoint-Higgs theory is S${}_c$-confining.

     What about the Higgs phase, in which center symmetry is spontaneously broken?  Here we can refer to ref.\  
\cite{Greensite:2004ke}, where the remnant symmetry order parameter $|u|$, defined in \rf{ut}, was computed numerically in both the confinement and Higgs phases of the gauge-adjoint Higgs theory \rf{ahiggs}. There it was found that $\langle |u| \rangle = 0$ in the center symmetric phase, and is non-zero
in the Higgs phase.  We know that $\langle |u| \rangle > 0$ implies a loss of S${}_c$-confinement, so we conclude that S${}_c$-confinement exists
in the center-symmetric phase, but not in the Higgs phase.

\subsection{Center symmetry}

    It has been suggested (see, e.g., chapter 3 of \cite{Greensite:2011zz}) that confinement might be identified with ``magnetic disorder,''
i.e.\ confinement describes a theory in which there exist disordering vacuum fluctuations strong enough to induce an area law falloff for Wilson loops at arbitrarily large scales.  By this definition, confinement in an SU(N) gauge theory is the phase of unbroken center symmetry, since a theory in which the center symmetry is broken either spontaneously (matter in the adjoint representation) or explicitly (matter in the fundamental representation) or was trivial from the beginning ($G_2$ gauge theory) are theories in which large planar Wilson loops have only a perimeter-law falloff.  The terminology has a price, however, since by this definition QCD is not a confining theory; it is only ``confinement-like'' at intermediate distance scales.   The center symmetry confinement criterion is much stronger than what we here call C-confinement, but like C-confinement it fails to clearly distinguish between theories with linear Regge trajectories and a spectrum of metastable flux tubes, such as QCD, and theories with only Yukawa forces, such as gauge-Higgs theory in the Higgs region.  The S${}_c$-confinement criterion advocated here does make this distinction.   

    Center symmetry is associated with the center vortex theory of confinement.  In recent years the evidence that center vortices underlie the non-perturbative behavior of gauge theories has become very strong.  Older work on this subject, reviewed in \cite{Greensite:2003bk,Greensite:2011zz}, was already quite persuasive, at least in our opinion, but with the recent work of Kamleh, Leinweber, and Trewartha \cite{Kamleh:2017lij,*Trewartha:2015ida,*Trewartha:2015nna,*Trewartha:2017ive} the vortex mechanism now seems very compelling.  To summarize the evidence in a sentence:  if one extracts the center vortices from an SU(3) configuration generated by lattice Monte Carlo, and subjects these vortices to a smoothing procedure, the resulting configurations are virtually identical, in their non-perturbative properties, to the unmodified configurations subjected to an equal amount of smoothing.  These properties include the string tension, chiral symmetry breaking, instanton density, and even, at least at the qualitative level, the particle spectrum.  

    While center symmetry is broken explicitly in the gauge + matter theories under consideration, this does not mean that center vortices are irrelevant when matter is added, and the vortex mechanism in gauge-Higgs theory has been investigated in a number of studies \cite{Greensite:2006ng,*Greensite:2004ur,*Bertle:2003pj}.  It is found that vortex removal also removes the linear part of the static quark potential, which exists in the confinement region prior to string breaking, while center projection, which is the mapping of lattice configurations onto a set of thin center vortices, preserves the string-breaking property.  Although the center vortex mechanism leads to an asymptotic area law falloff if the positions where vortices pierce a planar loop are uncorrelated at long distances, this lack of correlation need not (and presumably does not) apply in a gauge-matter theory.

\begin{figure}[htb]
\subfigure[~confinement phase]  
{   
 \label{v83}
 \includegraphics[scale=0.6]{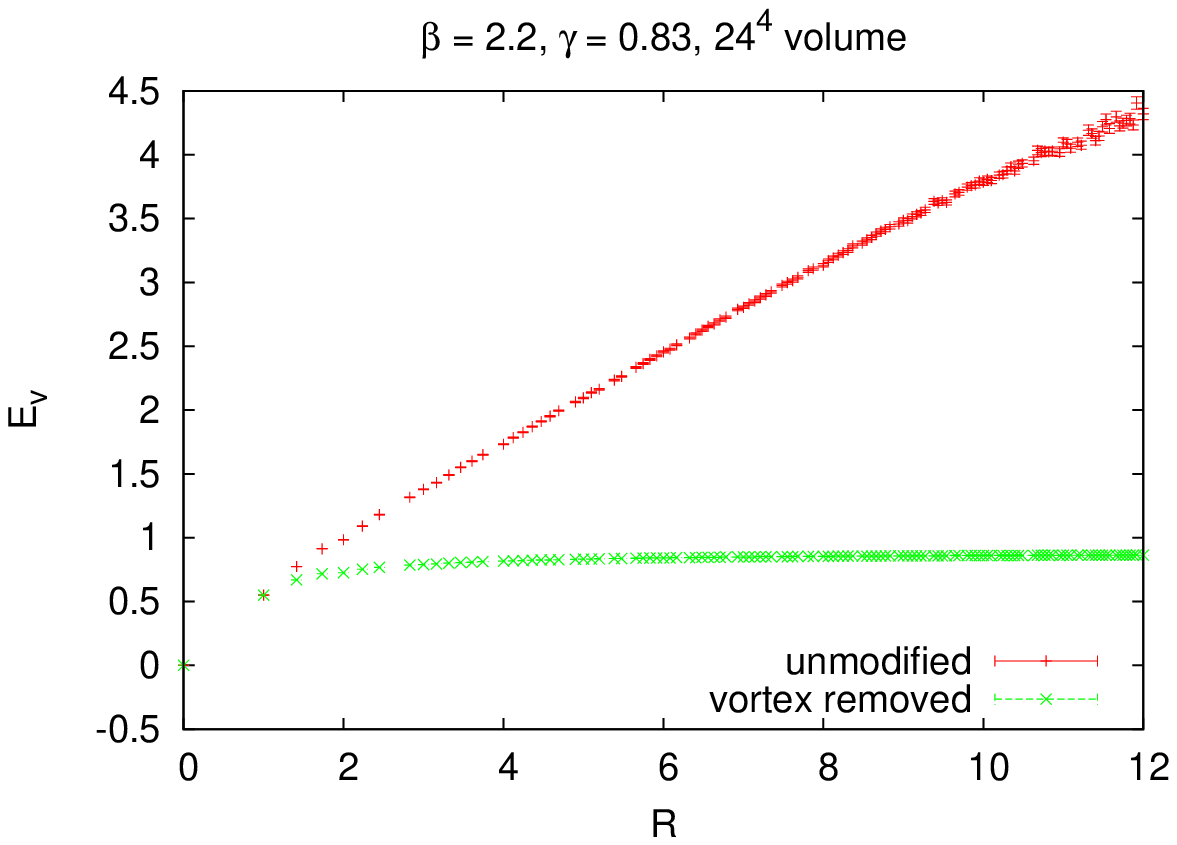}
}
\subfigure[~Higgs phase]
{   
 \label{v85}
 \includegraphics[scale=0.6]{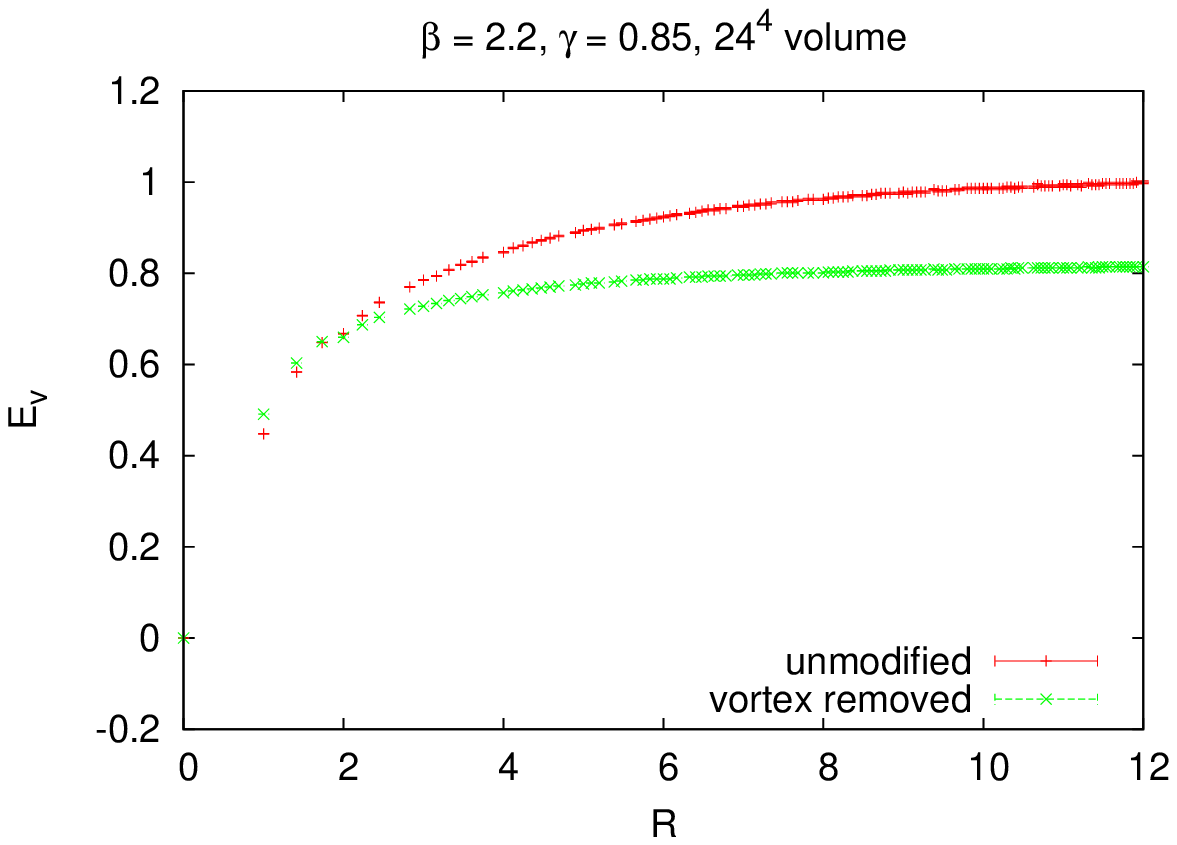}
}
\caption{The effect of vortex removal on $E_V(R)$ of \rf{Ecoul} at $\b=2.2$ on a $24^4$ lattice volume  in (a) the confinement region, at $\g=0.83$; and in (b) the Higgs region, at $\g=0.85$.  Both $\g$ values are in the immediate neighborhood of the confinement-Higgs transition at $\g=0.84$.}
\label{center}
\end{figure} 

    It is interesting to see what happens to the energy $E_V(R)$ defined in \rf{Ecoul}, when vortices are removed from
thermalized lattice configurations using the methods reviewed in, e.g., \cite{Greensite:2003bk}.  The effect in the
immediate neighborhood of the remnant-symmetry transition (here coinciding with a sharp thermodynamic crossover) at $\b=2.2, \g=0.84$, is shown in Fig.\ \ref{center}.  The effect of vortex removal on $E_V(R)$ in the confinement regime ($\g=0.83$), Fig.\ \ref{v83}, is drastic, converting a linearly rising energy expectation value to an asymptotically constant behavior.  The effect in the Higgs regime ($\g=0.85$) is much milder.  There is some effect due to vortex removal, as seen in Fig.\ \ref{v85}, but $E_V(R)$
levels off to a constant in both the unmodified and vortex removed situations.

\section{Finite temperature}

     If the observable $E_V(R)$ defined in eq.\ \rf{Ecoul} violates the S${}_c$-confinement bound, then the system is not S${}_c$-confining.  On the other hand, if $E_V(R)$ obeys the S${}_c$-confinement bound, it does not necessarily follow that the system is S${}_c$-confining, and the deconfined phase of a pure gauge theory is a case in point.
     
     Let $\{E_n\}$ denote the set of all energy eigenstates of a pure gauge theory, $\{E'_n(R)\}$ denote the corresponding set of energy eigenstates in the presence of a static quark-antiquark pair separated by a distance $R$.  Let $\b$ be the inverse temperature (not to be confused with lattice coupling) and let
\bea
           U(\b) &=& - {d\over d\b} \log \left[\sum_n e^{-\b  E_n} \right]  \non \\
                    &=&  \e(\b) \V_3 + \E_{vac} \non \\
           U'(\b,R) &=& - {d\over d\b} \log \left[\sum_n e^{-\b  E'_n(R)} \right]   \non \\
\label{U}
\eea
be the energies of the pure gauge system, and the gauge system containing static quarks respectively, at finite temperature. 
Here $\V_3$ is the spatial volume and $\e(\b)$ the finite temperature energy density above the ground state energy $\E_{vac}$. We
may write the energy of the quark-antiquark system as
\bea
          \E(\b,R)  &=& U'(\b,R) - U(\b)   \non \\
                    &=& - {d\over d\b} \log \left[{\sum_n e^{-\b  E'_n(R)} \over \sum_n e^{-\b  E_n} } \right] \non \\
                    &=& - {d\over d\b} \log \langle P(0) P^\dg(R) \rangle \ ,
\eea
where $\langle P(0) P^\dg(R) \rangle$ is the Polyakov line correlator.

    Above the deconfinement temperature $T_c=1/\b_c$, the $R$-dependence of $\E(\b,R)$ falls off exponentially with $R$.   This fact does not mean that one can find a physical state in the pure gauge theory which violates the S${}_c$-confinement bound \rf{criterion}, because for typical energy eigenstates contributing to the canonical ensemble at high temperature, the energy above the vacuum energy contains a contribution $\e(\b) \V_3$ proportional to the spatial volume. Because of the finite energy density $\e(\b)$,  the energy of these ``deconfined'' states will still bounded from below by $E_0(R)$, at least in the large volume limit.  

     At finite temperature, the S${}_c$-confinement criterion must be modified slightly.  We define
\bea
          Q_V &=&   \overline{q}^a(\vx) V^{ab}(\vx,\vy;A) q^b(\vy) \ ,  \non \\
          \Psi_{Vn} &=&  Q_V \Psi_n \ ,
\label{Q}
\eea
where, as before, $V^{ab}(\vx,\vy;A)$ is a bi-covariant operator depending only on the gauge field $A$, and the $\Psi_n$ are the 
energy eigenstates in the absence of the static quark sources.  Then the relevant energy above the thermal background
is
\bea
          \overline{E}_V(R) &=& { \sum_n \langle \Psi_{Vn}|H|\Psi_{Vn} \rangle e^{-\b E_n} \over
                                                 \sum_n  e^{-\b E_n} }  - U(\b) \non \\
                                       &=& {\tr[ (Q^\dg_V H Q_V) e^{-\b H}] \over \tr[ e^{-\b H}] } - U(\b) \ ,
\eea
where $H$ is the Hamiltonian in some physical gauge.  A gauge or gauge + matter system is S${}_c$-confining iff, for any bicovariant
operator  $V(\vx,\vy;A)$, the interaction energy $\overline{E}_V(R)$ is bounded from below by some asymptotically linear potential.
For a pure gauge theory this criterion is certainly violated in the deconfined phase, because the vast majority of energy eigenstates
containing static quarks, which contribute substantially in the canonical ensemble leading to $U'(\b,R)$ in \rf{U}, violate the bound, and because these states depend only on $A$ (and not on matter fields) they should belong to the class of states of the form \rf{Q}. 

    For QCD at finite temperature the situation is similar, in certain respects, to gauge-Higgs theory, in that there is no thermodynamic transition from the confinement phase to the quark-gluon plasma phase.  The phases are continuously connected under variation with temperature, and the question is how to properly distinguish between confinement and deconfinement in this case.  The S${}_c$-confinement criterion provides a possible answer.  We have argued that there are good reasons to believe that QCD, like pure SU(N) gauge theory, is an S${}_c$-confining theory, at least at low temperatures.  It is certain that this property is lost at high temperatures in the case of pure gauge theories, because the energy eigenstates containing static quarks, and which dominate the canonical ensemble, violate the S${}_c$-confinement bound.   We cannot be equally certain of the loss of S${}_c$-confinement in high temperature QCD, because the energy eigenstates containing static quarks also depend on the light quark fields, and it is unlikely that these states have the form \rf{Q}, with the operator $V$ independent of the light quark fields.  So the loss of 
S${}_c$-confinement in the quark-gluon plasma is a conjecture, but we believe it is a very reasonable conjecture, since we see no reason that 
S${}_c$-confinement should be lost in the high temperature  phase of a pure gauge theory, yet persist at high temperatures in a gauge + matter theory.  

    To verify this conjecture numerically, it is necessary to find some bi-covariant operator $V$ which violates the S${}_c$-confinement
criterion at high temperature.   The operator in eq.\ \rf{VCoul}, used to demonstrate the absence of S${}_c$-confinement in the Higgs
region of gauge-Higgs theory, is a natural candidate.  But, a little surprisingly, this operator fails to detect the loss of S${}_c$-confinement
at high temperatures, even in the pure gauge theory.  This was shown in ref.\ \cite{Greensite:2004ke}.  We have also checked the
pseudo-matter operator \rf{pseudo} and the fat link operator \rf{Vfat}, but here again we do not see a loss of S${}_c$-confinement across the transition.  We are certain that
the deconfined phase is not S${}_c$-confining in the pure gauge theory, for the reasons stated above, and believe this is also very likely in 
gauge + matter theories.  But the actual construction of an
operator which reveals the loss of S${}_c$-confinement in the quark-gluon plasma, thereby demonstrating a sharp separation between
the confinement and high temperature phases of QCD, is at present an open problem.

\section{Conclusions}

   We have proposed a generalization of the Wilson area law criterion, namely charge-separation confinement (or ``S${}_c$-confinement''),  
as a criterion for confinement in gauge theories with matter fields transforming in the fundamental representation of the gauge group.
Like the area law, S${}_c$-confinement is much stronger than the condition that asymptotic particle states are color singlets (``C-confinement''), which holds even for gauge-Higgs theories in the Higgs regime.   Under this new criterion, if we consider the subclass of physical states
\rf{Vstate} containing a static quark-antiquark pair whose color charges are unscreened by matter fields, then the theory is S${}_c$-confining if
the energy of such states is bounded from below by a linear potential.  In theories with a non-trivial global center symmetry, 
S${}_c$-confinement is equivalent to the area-law criterion, and therefore holds in all gauge theories with unbroken center symmetry.  In this article we have conjectured that certain theories in which center symmetry is broken explicitly by matter fields, such as QCD and gauge-Higgs theory, are also S${}_c$-confining (at least, in the latter case, in some region of the phase diagram).   We have illustrated this conjecture numerically by applying certain operators (Dirac and pseudo-matter) to
the vacuum of a gauge-Higgs theory, and showing that the resulting physical states distinguish between a C-confining phase, and a phase which may be S${}_c$-confining.

   It is understood, of course, that illustrations are not a proof.  We believe that the S${}_c$-confinement conjecture for QCD is reasonable,
and it is supported by the existence of linear Regge trajectories in the spectrum of resonances.  But if it should turn out that QCD and other gauge + matter theories are not S${}_c$-confining, then we must return to the view that there is no rigorous way of distinguishing between, e.g., the Higgs and confining phases of a gauge theory, and that S${}_c$-confinement is a property that applies only to gauge theories with an unbroken center symmetry.

    Apart from the question of proof, which is unlikely to be supplied in the near future, there are other open questions.  First, we would like
to construct a $V$ operator which shows the breakdown of S${}_c$-confinement in the high-temperature deconfined phase; the Dirac and 
pseudo-matter fields do not seem adequate for this purpose.  Also, the present definition of S${}_c$-confinement requires that the gauge group
has a non-trivial center, so that there exist matter fields which cannot be trivially screened by gluons.  This is a limitation, because we
cannot so far apply the criterion to, e.g., $G_2$ or SU(N)/$Z_N$ gauge theories, and it would be interesting to see if there is a way to
extend the S${}_c$-confinement criterion to those theories as well.  We reserve these questions for future investigation.

\bibliography{scon}

\end{document}